\documentclass{article}
\usepackage{graphicx}
\usepackage{mathrsfs}
\usepackage{amsfonts}
\usepackage{amsmath}
\usepackage{amsthm}
\usepackage{amssymb}
\usepackage{ragged2e}
\usepackage{fancyhdr}
\usepackage{comment}
\usepackage{float}
\usepackage{url}
\usepackage{bm}
\usepackage{optidef}
\usepackage{subcaption}
\usepackage[a4paper, portrait, margin=1in]{geometry}
\usepackage[style=apa]{biblatex}
\usepackage{multirow}
\usepackage[title, toc, page, titletoc]{appendix}
\usepackage{lastpage}
\usepackage[colorlinks=true,linkcolor=blue,citecolor=blue,urlcolor=blue]{hyperref}
\usepackage{listings}
\usepackage{color}
\usepackage{booktabs}
\usepackage{parskip}
\usepackage[ruled,vlined]{algorithm2e}

\hypersetup{pdfborder=0 0 0}

\newcommand\norm[1]{\left\lVert#1\right\rVert}
\newcommand\codify[1]{\textcolor{blue}{\texttt{#1}}}

\definecolor{dkgreen}{rgb}{0,0.6,0}
\definecolor{gray}{rgb}{0.5,0.5,0.5}
\definecolor{mauve}{rgb}{0.58,0,0.82}

\providecommand{\keywords}[1]
{
  \small	
  \textbf{Keywords:} #1
}

\makeatletter
\renewcommand\@date{{%
  \vspace{-2\baselineskip}%

  {\large\centering
  \begin{tabular}{@{}c@{}}
    Justin Philip Tuazon\textsuperscript{1} \\
  \end{tabular}%
  \quad\quad
  \begin{tabular}{@{}c@{}}
    Gia Mizrane Abubo \\
  \end{tabular}
  \quad\quad
  \begin{tabular}{@{}c@{}}
    Joemari Olea\textsuperscript{2} \\
  \end{tabular}}
  
  \smallskip

  {\normalsize\textsuperscript{1}Department of Computer Science, University of the Philippines Diliman\linebreak
  \textsuperscript{2}Department of Educational Psychology, University of Texas at Austin}

  \smallskip

  {\large November 2026}
}}
\makeatother

\makeatletter
\def\@maketitle{%
  \newpage
  \null
  \vskip 2em%
  \begin{center}%
  \let \footnote \thanks
    {\LARGE\bfseries\@title \par}%
    \vskip 1.5em%
    {\large
      \lineskip .5em%
      \begin{tabular}[t]{c}%
        \@author
      \end{tabular}\par}%
    \vskip 1em%
    {\large \@date}%
  \end{center}%
  \par
  \vskip 1.5em}
\makeatother

\fancypagestyle{firststyle}
{
   \fancyhf{}
   \fancyfoot[R]{\small Page \thepage\ of \pageref{LastPage}}
   \fancyfoot[L]{\small Preprint}
   
}

\title{Pairwise Target Rotation for Factor Models}
\author{}

\begin{document}
\justifying
\maketitle
\thispagestyle{firststyle}
\begin{abstract}
    \noindent Factor analysis is used to characterize latent variables by examining relationships among manifest variables. In exploratory factor analysis (EFA), various factor models are considered to uncover the latent structure. Now, the success of EFA lies with the model's interpretability, as the objective is to build a factor model that is not only supported by data, but is also meaningful. Achieving such, however, is challenging, as gauging interpretability is difficult and subjective, owing to rotational indeterminacy. Thus, we propose a new index that measures the interpretability of a factor model. The index does this by evaluating the agreement between a priori information, such as semantic information from items, and the loadings. We also introduce pairwise target rotation, a rotation method that maximizes interpretability based on the index. In general, this method allows for an intuitive yet flexible way of incorporating a priori information, such as semantics, in factor rotations, which can help in performing EFA more effectively. Based on simulations, the index correctly indicates better fit when noise levels are lower and the new rotation outperforms classical orthogonal rotations across diverse conditions in terms of recovering the latent structure. Finally, we applied the method to empirical datasets, providing evidence linking item semantics and latent constructs, consistent with recent findings. 
\end{abstract}

\keywords{factor analysis, factor models, interpretability, latent variables, natural language processing}

\tableofcontents

\section{Introduction}\label{sec:1}
Factor analysis can be understood in several ways. One popular way is that it is a method for characterizing the relationships among many manifest or observed variables in terms of a smaller number of latent variables called \textit{factors} \parencite{stat147Ref,Mulaik2010}. Under this perspective, variables are ``grouped" based on their correlations in such a way that variables in the same group are highly correlated with one another, while variables that belong to different groups are only weakly correlated. Meanwhile, factor analysis can be also be viewed as a method for identifying underlying factors, typically conceptualized as continuous variables, that account for the shared variation among measured observed variables \parencite{bartholomew2011latent,brown2016confirmatory}.

Since its development \parencite{spearman1904general,spearman1927abilities}, factor analysis has become an important method in psychometrics and related fields. It is commonly used to identify latent constructs, examine the dimensionality of an instrument, support scale development, provide evidence of construct validity, reduce the dimensionality of a dataset, and estimate factor scores representing individuals' standings on the constructs being measured \parencite{bartholomew2011latent,brown2016confirmatory,furr2021psychometrics,stat147Ref,kline2016principles,Mulaik2010}. As a general example, factor analysis has historically been used to study intelligence by representing performance across several cognitive tasks in terms of underlying latent abilities \parencite{stat147Ref,Mulaik2010}. Beyond intelligence research, factor analysis has been widely applied in several other disciplines. For instance, the method has been used to evaluate measures of sustainable agricultural development \parencite{laurett2021} and to study study latent dimensions relevant to transportation and traffic \parencite{ledesma2021}.

Ultimately, factor analysis is not limited to a single domain; rather, it provides a general framework for studying latent constructs and as discussed in \textcite{enhancedInterpretability2020}, has become a useful tool for the broad task of extracting knowledge, regardless of the area. In this paper, we focus on \textit{exploratory factor analysis} or EFA (in contrast to confirmatory factor analysis or CFA), which is used to explore underlying latent constructs and includes methods for identifying and estimating factors, and finding interpretable factor solutions. In particular, we propose a new criterion, as well as a factor rotation method based on it, to help address the critical challenge of producing interpretable factor solutions in the implementation of EFA. 

As noted by \textcite{enhancedInterpretability2020}, researchers who employ EFA face the non-trivial task of interpreting the extracted factors. Crucially, the substantive value of factor analysis depends heavily on whether the resulting factors can be interpreted as meaningful latent constructs. This concern is reflected in what \textcite{stat147Ref} refer to as the ``WOW criterion", where the success of an application of factor analysis depends on whether the factor model is interpretable or not. In more concrete terms, a factor solution is useful when the extracted factors provide a coherent and theoretically meaningful explanation of the observed relationships among manifest variables (i.e., when the factor solution is interpretable). That is to say, if the extracted factors are weakly justified or interpreted inconsistently with their indicators, then the resulting claims about the latent constructs may be unsupported \parencite{knekta2019one,messick1995validity,tavakol2020factor}.

Unfortunately, interpretability, which is the extent to which a factor solution supports a coherent substantive explanation of the latent constructs, is difficult to assess and quantify because it often largely depends on the researcher's judgment, substantive context, and theoretical expectations. The ``interpretability problem" (i.e., the challenge of obtaining and deciding on an interpretable factor solution) is further complicated by the well-known fact that the factor model is \textit{rotationally indeterminate}, such that different rotated solutions (i.e., rotated versions of the loading matrix) can fit the data similarly well (i.e., produce the same model-implied covariance matrix for the manifest variables) despite yielding different substantive interpretations \parencite{gorsuch1983,fabrigar1999evaluating,costello2005best,Mulaik2010}. To put it simply, the interpretability problem is compounded by the fact that the model is \textit{not identifiable}.

To address the interpretability problem, rotation methods are commonly used to obtain a loading matrix with a more interpretable structure \parencite{enhancedInterpretability2020}. These include sparsity-based rotation methods, methods that use auxiliary or explanatory variables, and methods that use \textit{a priori} information. Because our proposed method is concerned with the interpretability problem, we briefly summarize below several major approaches that have been used to make factor solutions more meaningful.

Sparsity-based approaches aim to improve interpretability by producing loading matrices in which most loadings are near zero and only a few are large \parencite{Browne2001MBR}. In a sparse loading matrix, each manifest variable is strongly associated with only a few factors, and each factor is defined by a limited set of manifest variables. This aligns with the concept of simple structure, concretized in the perfect cluster configuration discussed by \textcite{Browne2001MBR} and in Thurstone's rules for simple structure from \textcite{thurstone1947} \parencite[as cited in][]{Browne2001MBR}, and is encouraged by classical rotation methods, such as varimax \parencite{Kaiser1958} and the general orthomax family of rotations \parencite{jennrich2003}, and by various Bayesian techniques \parencite{bayesianKaufmann2017,yamamoto2017,bayesianZhao2016}. However, aiming for sparsity alone does not ensure that factors are substantially meaningful and can even produce poor model fit to the data, such as in cases where regularization is introduced to achieve sparsity \parencite{enhancedInterpretability2020}. For example, a sparse loading matrix may still be difficult to interpret if the variables that load on the same factor do not share a coherent substantive meaning. Indeed, a factor may be defined by a small number of variables, but if those variables refer to different psychological, behavioral, or substantive domains, then the resulting factor may not have a clear interpretation.

Another general approach is to use information from ``external" variables beyond the loading matrix. In some settings, the researchers may have access to auxiliary or explanatory variables that are not necessarily part of the original set of manifest variables but are expected to be related to the latent factors. The general idea is that the relationship between these auxiliary or explanatory variables and the factors can provide additional evidence for interpreting the factors \parencite{fan2016,liJung2017,Taeb2017,Taeb2018}. Nevertheless, this approach does not necessarily provide a complete interpretation of the factors. For instance, an auxiliary variable may be associated with a factor without fully explaining the factor's substantive meaning. In addition, challenges similar to those faced when using the original set of manifest variables may persist, such as in cases where a single factor is associated with several (possibly incoherent) auxiliary variables. Ultimately, auxiliary or explanatory variables may end up just introducing additional complexity as they may only partially explain factors \parencite{enhancedInterpretability2020}, and the use of such do not remove the need to examine the original set of manifest variables anyway. An adjacent method from \textcite{enhancedInterpretability2020}, which directly matches explanatory variables to factors through optimization, simplifies the use of explanatory variables and can potentially aid interpretation as well, but may also face problems similar to those described earlier. Moreover, it uses estimated factor scores and as such, suffers from the \textit{factor score indeterminacy} problem discussed in \textcite{Mulaik2010}.

Finally, the last important class of methods that we discuss here incorporates \textit{a priori} information when rotating the loading matrix or factor solution. These methods take advantage of rotational indeterminacy by choosing, among the many equivalent rotated solutions, the one that best agrees with a prespecified structure. In this sense, rotational indeterminacy is not only a problem but is also an opportunity. Because multiple rotated loading matrices can reproduce the same covariance structure, the researcher can select one that is more substantively meaningful. The primary method under this class is \textit{target rotation} \parencite{targetOrth1972,targetObli1972,target2018}. The loading matrix is rotated to match, as closely as possible, specified loading values and if desired, even factor correlations. However, target rotation can be difficult to use in exploratory settings, where the researcher may not have adequately strong hypotheses about the factor structure and/or may not have prior expectations that can be easily expressed as precise loadings. For instance, the researcher may have only qualitative expectations such that they believe two items are conceptually similar without knowing exact numerical loading values. Moreover, rotated solutions may be sensitive to the particular values chosen by the researcher. The regularization approach from \textcite{wu2025} is useful and can be considered as an adjacent method. There, one can specify the general desired structure $\mathscr{S}$ (e.g., simple structure, bi-factor) instead of specific loadings, such that the matrix $S\in\mathscr{S}$ if and only if $S$ follows the desired structure. Although, expectations about the latent structure itself are sometimes not available and further prior information beyond the desired structure may need to be specified (e.g., the regularization strength, such as the loading threshold $\rho$ in the penalty term $\rho\lVert S\rVert_{0}$).

Fundamentally, target rotation and existing adjacent methods provide powerful frameworks for incorporating prior information, but they also illustrate a broader limitation: many forms of prior information are not naturally expressed as numerical loading targets or specific factor structures. In psychometric applications, prior information often comes from the wording or meaning of items, theoretical expectations about constructs, or expert judgment about item similarity. These forms of information may be highly relevant to interpretation but difficult to encode in a traditional target matrix.

Therefore, to provide an alternative way of incorporating prior information into factor rotation, we propose a variant of target roation for exploratory factor analysis called \textit{priorimax} or pairwise target rotation. Instead of requiring a target loading matrix, priorimax uses pairwise similarities among manifest variables. These similarities may be based on semantic content, expert judgment, theoretical expectations, or any other source of prior information that can be represented as a similarity matrix. That is, in our approach, a similarity matrix of the manifest variables, called the \textit{prior matrix}, is specified or calculated before rotation. The information from the prior matrix is then used to guide the rotated solution. Particularly, the proposed method evaluates the agreement between the loading matrix and the prior matrix using an interpretability index. Higher values of this index indicate that the (rotated) loading matrix is more consistent with the \textit{a priori} information. Ultimately, the priorimax rotation optimizes this criterion over the space of possible rotations.

This approach differs from classical rotation methods because it does not primarily seek simple structure or sparsity, which is made clear by relating it to the orthomax family of rotations. Instead, it seeks a rotated loading matrix whose structure agrees with prior information about the relationships among the manifest variables. It also differs from standard target rotation and the regularization approach because the prior information is not expressed as a target loading matrix. Rather, the prior information is expressed as pairwise similarities among manifest variables. This may be more intuitive in exploratory settings, where researchers often possess useful \textit{a priori}, but usually qualitative and/or incomplete, information, such as semantic relationships among items or expert judgments, that cannot be represented as exact numerical target values. For instance, it may be easier for the researcher to specify which items are similar to one another than to specify exact loadings. Finally, because the interpretability index is computed from the loading matrix and the prior matrix, the method does not require additional auxiliary variables and does not utilize estimated factor scores, avoiding the factor score indeterminacy problem.

Based on our simulation study, we found that our proposed index is indeed a suitable goodness-of-fit measure, as it correctly indicates a better fit when noise levels are lower. Furthermore, we found that our proposed rotation method, the priorimax rotation, performs better compared to other orthogonal rotation methods in terms of recovering the true structure, provided that the prior information (e.g., semantic information) is a good enough proxy for the loadings. Its advantages are particularly pronounced when cross-loadings are present. In fact, in such cases, simulations suggest that the priorimax rotation with the sample correlation matrix (of the manifest variables) as the prior matrix yields better recovery of the true structure compared to classical orthogonal rotations. This means that error can be reduced without introducing any new data or information through the simple (re-)use of the sample correlation matrix in factor rotation. At the same time, the empirical case studies further substantiate the relationship between semantics and psychological constructs, as we were able to demonstrate that a high interpretability index $\omega$ can be empirically achieved through the priorimax rotation. These results also highlight the effectiveness of the rotation method in producing interpretable factor solutions.

The rest of the paper is structured as follows. Section \ref{sec:2} provides a review of the common factor model. Meanwhile, in Section \ref{sec:3}, we detail our proposed method for measuring ``interpretability" and obtaining ``interpretable" factor models. In this section, we also define what ``interpretable" means in the context of factor solutions, as well as provide the premise and motivation for the definition. In Section \ref{sec:4}, we demonstrate and further characterize our proposed method in two ways: 1) through a simulation study and 2) through an empirical case study. Finally, we end with Section \ref{sec:5}, where we summarize the contributions of this paper and suggest directions for future related research.

\section{Review of The Common Factor Model}\label{sec:2}
We first introduce the common factor model. 
Suppose that there are $M\in\mathbb{Z}^{+}$ manifest variables denoted by $X_{1},X_{2},\dots,X_{M}$ and that there are $T\in\mathbb{Z}^{+}$ common factors (i.e., latent variables) denoted by $F_{1},F_{2},\dots,F_{T}$, with $T<M$ (ideally, $T\ll M$). As mentioned in \textcite{stat147Ref}, the factor model $\mathscr{F}$ is
$$X_{i}=\mu_{i}+\sum_{k=1}^{T}{l_{i,k}F_{k}}+\varepsilon_{i},$$
where $\mu_{i}=\mathbb{E}\left(X_{i}\right)$, $l_{i,k}$ is the \textit{loading} of manifest variable $X_{i}$ on factor $F_{k}$, and $\varepsilon_{i}$ is a random error term. The loading $l_{i,k}$ describes the strength and direction of the relationship between the manifest variable $X_i$ and the common factor $F_K$. When the variables and factors are standardized, a large positive (negative) loading suggests that higher values on $F_k$ are associated with higher (lower) values of $X_i$. On the other hand, loadings close to zero suggest that $X_i$ is only weakly related to that factor. Thus, in applications, the loading matrix is used to determine which manifest variables are most strongly associated with each factor and, therefore, to help interpret the substantive meaning of the factors. More compactly,
\begin{equation}
\bm{\underline{X}}_{M\times1}=\bm{\underline{\mu}}_{M\times1}+\bm{\underline{L}}_{M\times T}\bm{\underline{F}}_{T\times 1}+\bm{\underline{\varepsilon}}_{M\times1},
\end{equation}
where
$$
\bm{\underline{X}}=\begin{bmatrix}
  X_{1} \\
  X_{2} \\
  \vdots \\
  X_{M}
\end{bmatrix}, ~
\bm{\underline{\mu}}=\begin{bmatrix}
  \mu_{1} \\
  \mu_{2} \\
  \vdots \\
  \mu_{M}
\end{bmatrix}, ~
\bm{\underline{L}}=\begin{bmatrix}
  l_{1,1} & \dots & l_{1,T} \\
  \vdots & \ddots & \vdots \\
  l_{M,1} & \dots & l_{M,T}
\end{bmatrix}, ~
\bm{\underline{F}}=\begin{bmatrix}
  F_{1} \\
  F_{2} \\
  \vdots \\
  F_{M}
\end{bmatrix}, ~\text{and}~
\bm{\underline{\varepsilon}}=\begin{bmatrix}
  \varepsilon_{1} \\
  \varepsilon_{2} \\
  \vdots \\
  \varepsilon_{M}
\end{bmatrix}
$$

Lastly, several model assumptions are made:
\begin{enumerate}
    \item $\bm{\underline{F}}$ and $\bm{\underline{\varepsilon}}$ are stochastically independent.
    \item $\mathbb{E}\left(\bm{\underline{F}}\right)=\bm{\underline{0}}$ and $\mathbb{E}\left(\bm{\underline{\varepsilon}}\right)=\bm{\underline{0}}$.
    \item $\text{Cov}\left(\bm{\underline{F}}\right)=\bm{\underline{I}}$ and $\text{Cov}\left(\bm{\underline{\varepsilon}}\right)=\bm{\underline{\Psi}}$ for some diagonal matrix $\bm{\underline{\Psi}}$.
\end{enumerate}
Collectively, these assumptions make $\mathscr{F}$ an \textit{orthogonal factor model} \parencite{stat147Ref}. Relaxing the assumption that $\text{Cov}\left(\bm{\underline{F}}\right)=\bm{\underline{I}}$ by allowing the covariances among factors to be possibly non-zero results to the \textit{oblique factor model}. Nonetheless, we focus primarily on the orthogonal factor model in this paper.

Although factor analysis can be used for several purposes, a central task in many applications is to assign substantive meaning to the extracted factors. This task is non-trivial because the factors are not directly observed; rather, they are inferred from patterns of association among the manifest variables. Thus, after a factor model is estimated, researchers must determine whether the resulting latent dimensions correspond to meaningful psychological, educational, behavioral, or substantive constructs \parencite{enhancedInterpretability2020}.

To interpret the factors, researchers typically examine the loading matrix $\bm{\underline{L}}$. In practice, this means identifying which manifest variables have large loadings on each factor and then using the content or meaning of those variables to assign a substantive label to the factor. For example, if several questionnaire items about anxiety have large positive loadings on the same factor, the researcher may interpret that factor as an anxiety-related dimension. Conversely, if a factor has large loadings from variables with no clear substantive connection, the factor may be difficult to interpret. Researchers also examine whether each manifest variable loads strongly on only one factor or on multiple factors. A factor solution is usually easier to interpret when each factor is associated with a distinct subset of manifest variables and when cross-loadings are small.

However, as mentioned earlier, factor models are generally affected by rotational indeterminacy. That is, there may be multiple mathematically equivalent loading matrices that reproduce the same covariance or correlation structure among the manifest variables. For example, in the orthogonal factor model, replacing the loading matrix $\bm{\underline{L}}$ with
$$
\bm{\underline{L}}^{*}=\bm{\underline{L}}\bm{\underline{R}},
$$
where $\bm{\underline{R}}$ is an orthogonal rotation matrix, does not change the model-implied covariance or correlation matrix of the manifest variables, and the assumptions of the orthogonal factor model continue to hold \parencite{Mulaik2010}. Thus, the same observed data can correspond to different rotated loading matrices. This property is not merely a technical issue; it directly affects interpretation because different rotations can suggest different substantive meanings for the factors. Consequently, rotation methods are often used to select, among many equivalent representations, a loading matrix that is easier to interpret \parencite{enhancedInterpretability2020}. 

\section{Proposed Method}\label{sec:3}

\subsection{Premise for Interpretability}
The preceding discussion shows that existing approaches to factor interpretability generally rely on one of three sources of information: the numerical pattern of the loading matrix, external variables associated with the latent factors, or prior expectations imposed through rotation. The proposed method builds from these traditions by treating the semantic content of the manifest variables as another form of \textit{a priori} information. In particular, when manifest variables are associated with questionnaire or test items, the items' wording provide information about the constructs they are intended to measure. The goal of the proposed method is therefore to use this semantic information to guide the choice of rotation and to favor loading matrices in which variables with similar semantic meanings have similar loading profiles. Note, however, that while we first assume that manifest variables have associated questions to more easily motivate our approach, we later relax this assumption without prejudice to the premise.

Now, because items are generated based on the target construct(s), it is natural to hypothesize that information about the construct is ``embedded" in the questions themselves, and perhaps these pieces of information can be used to identify psychometric factors (as we later incorporate in our proposed method). In fact, the Deep Lexical Hypothesis from \textcite{deepLexical2022} found that psychological constructs are fundamentally interlaced with the language used in the items. By comparing language models and survey data from the Big Five in \textcite{bigFive}, \textcite{deepLexical2022} found that responses themselves were reflected in the \textit{semantics} of the questions, such that the latent structure derived from natural language was similar to the one derived from actual responses.

Building on the Deep Lexical Hypothesis from \textcite{deepLexical2022}, \textcite{semanticLoadings} argued that there is a relationship between the semantic content of items and the item ratings, and that this relationship can be exploited to semantically characterize a psychometric factor (which represents a construct). To this effect, \textcite{semanticLoadings} introduced \textit{semantic loadings}. Roughly speaking, the ``words" (i.e., tokens) from the set of questions are ``clustered" into semantic communities (e.g., using one of many community detection algorithms), each of which is then interpreted. Separately, the psychometric factors are also identified (e.g., using factor analysis). Then, the semantic loading measures the overlap between a semantic community and a psychometric factor, which semantically characterizes the factor. Using the DASS questionnaire from \textcite{lovibond} as a case study, \textcite{semanticLoadings} found that semantic communities can identify aspects of constructs, as evidenced by statistically significant semantic loadings. This is consistent with the findings of \textcite{deepLexical2022}.

More recently, there have also been other studies establishing the link between item semantics and psychological constructs, consistent with the findings of the earlier works of \textcite{deepLexical2022} and \textcite{semanticLoadings}. These studies followed approaches similar to the strategy of \textcite{deepLexical2022}. For example, \textcite{pfa2025} proposed \textit{pseudo-factor analysis}, a data-less approach to factor analysis that attempts to identify the latent structure using solely semantic similarities among items. Using pseudo-factor analysis, \textcite{pfa2025} were able to recover the validated empirical factor structures of two personality frameworks, the Big Five IPIP-NEO-300 inventory from \textcite{ipipNEO} and the HEXACO model described in \textcite{hexaco}. Additionally, \textcite{milano2025} examined several questionnaires and found that items that belong to the same construct generally have higher semantic similarities compared to items that belong to different constructs. Moreover, they noted that the latent structure derived from item semantics is similar to the one derived from actual responses, arriving at the same conclusion as \textcite{deepLexical2022}.

Indeed, item semantics and responses are closely related, as established by these previous works involving various case studies. It is then on this note that we finally introduce the ``interpretability" criterion in our proposed method. In general, while previous methods use either only data from manifest variables (e.g., item responses) or only item semantics, as in the case of pseudo-factor analysis \parencite{pfa2025}, our approach uses both simultaneously to actively produce factor solutions that are both supported by data and interpretable. In particular, we exploit the relationship between semantics and responses, and leverage both semantic similarities and manifest variable correlations, to define what ``interpretable" means in the context of factor analysis. We then use this definition to actively influence the loading matrix (i.e., through a rotation matrix). In other words, whereas previous works focused on establishing a relationship between semantics and responses, we now rely on this relationship to define and incorporate an appropriate criterion for factor analysis.

Now, loosely speaking, a factor model is interpretable if the ``groupings" of manifest variables into factors makes sense based on semantics (i.e., based on the meanings of the questions). Particularly, questions that have similar meanings must also have similar responses. In other words, semantically similar questions should measure the same construct, so that items that indicate a factor are semantically ``coherent". For example, if the two questions ``I enjoy being in large social gatherings with many people." and ``I feel energized when I spend time at parties." are asked to the same respondent, the responses will most likely be similar, as both items roughly talk about ``socializing". Contrast this with a semantically dissimilar (or less semantically similar) question such as ``I find it easy to concentrate on complex problems for long periods of time.", which likely measures a different construct.

Operationalizing this, questions that are more \textit{semantically similar} should have corresponding manifest variables that are more \textit{loading-wise similar} (i.e., similar in terms of their loadings). Intuitively, as far as interpretability is concerned in factor analysis, if two manifest variables ``talk" about the same or similar things, then their variances must also be ``broken down" in similar (not necessarily exactly the same) ways. It is important to note that the premise here does not contradict the factor model or the model's assumptions. The relationships between manifest variables are still estimated or ``dictated" by the factor model and not directly by the language. Instead, interpretability is just defined to be the degree to which the interpretability criterion is satisfied, which primarily lies on the semantics or meanings of the questions (via semantic similarity).

While the estimation of the model (i.e., the latent structure) is through factor analysis, the interpretability criterion measures how interpretable the estimated model is. With this, semantic information from the questions (specifically, semantic similarities) can be used as \textit{a priori} information --- similar to how \textit{a priori} ``expectations" can be incorporated in target rotation as in \textcite{targetOrth1972}, \textcite{targetObli1972}, and \textcite{target2018} --- to influence the loading matrix and ultimately resolve the rotational indeterminacy of the factor model. More generally, the \textit{a priori} information or expectations from item semantics can be used to assess how interpretable the factor model is.

\subsection{Similarity Functions}
In our discussion of our premise for interpretability, we mentioned the terms ``semantically similar" and ``loading-wise similar". Here, we formally define \textit{semantic similarity} and \textit{loading similarity}, which will form the \textit{interpretability index} in the later parts.

First, let us define semantic similarity. Suppose that the question associated with the $k$th manifest variable $X_{k}$ is $Q_{k}$. Because each $Q_{k}$ is a sentence or statement, the questions can be represented (i.e., embedded or vectorized) as a vector of real numbers in some Euclidean space (i.e., converted into a useful numerical representation), as described in \textcite{nlp2018}. Using embeddings instead of the raw natural language text allows us to accomplish various natural language tasks that would otherwise be difficult or even impossible, such as precisely measuring semantic similarity. Usually, raw embeddings themselves are not useful, but functions \textit{defined} on the embeddings can be extremely valuable.

Nowadays, there are various foundational models (i.e., embedders) that can vectorize natural language input. Here, we use the \textit{universal sentence encoder} $G\left(\cdot\right)$ from \textcite{use:2018}, which maps arbitrary-length sentences into $\mathbb{R}^{512}$. Although, note that one can use any other model for embedding questions as the proposed method merely requires that we can calculate semantic similarities between pairs of sentences\footnote{In this case, we chose the universal sentence encoder as its embeddings were designed to be general-purpose. Although, the choice of embedder is arbitrary. In specialized cases, there may be models trained specifically on the topic of interest, which may be more appropriate.}. Given the embeddings from $G\left(\cdot\right)$, \textcite{use:2018} noted that the semantic similarity between any two questions $Q_{i}$ and $Q_{j}$ can be computed by the similarity function $D_{s}:\mathbb{R}^{512}\times\mathbb{R}^{512}\to\left[0,1\right]$ defined as
\begin{equation}
D_{s}\left(G\left(Q_{i}\right),G\left(Q_{j}\right)\right):=1-\frac{1}{\pi}\arccos{\frac{G\left(Q_{i}\right)\cdot G\left(Q_{j}\right)}{\norm{G\left(Q_{i}\right)}\norm{G\left(Q_{j}\right)}}}
\end{equation}

Note that $D_{s}$ is a function of the \textit{angular distance} (i.e., the angle formed by two vectors), mapped in a way such that $\text{ran}\left(D_{s}\right)=\left[0,1\right]$ and that higher values indicate higher degrees of semantic similarity. \textcite{use:2018} found that using a function based on the angular distance provides a better measure of semantic similarity compared to using raw cosine similarity.

Although there is no hard-and-fast rule for ``classifying" a pair of questions as either similar or not similar, practically speaking, the two questions are semantically the same if and only if $D_{s}=1$ and are semantically opposite if and only if $D_{s}=0$, with higher values of $D_{s}$ indicating more similarity in \textit{meaning}. In line with this, one can construct the \textit{questions semantic similarity matrix} $\bm{\bm{\underline{Q}}}=\left(D_{s}\left(G\left(Q_{i}\right),G\left(Q_{j}\right)\right)\right)_{1\leq i,j \leq M}$. Clearly, the diagonal entries of $\bm{\bm{\underline{Q}}}$ are all $1$. Moreover, $\bm{\bm{\underline{Q}}}$ is symmetric.

This time, let us define \textit{loading similarity}. For the factor model $\mathscr{F}$, recall that $\text{Cov}\left(\bm{\underline{X}},\bm{\underline{F}}\right)=\bm{\underline{L}}$ and that $\text{Cov}\left(\bm{\underline{X}}\right)=\bm{\underline{L}}\bm{\underline{L}}^{\prime}+\bm{\underline{\Psi}}$. Then,
$$
\text{Var}\left(X_{i}\right)=\sum_{j=1}^{T}{l_{i,j}^{2}}+\psi_{i}=h_{i}^{2}+\psi_{i},
$$
where $h_{i}^{2}$ is the \textit{communality} of $X_{i}$ and $\psi_{i}$ is the specific variance of $X_{i}$.

The communality $h_{i}^{2}$ is the part of $\text{Var}\left(X_{i}\right)$ that is due to or contributed by the $T$ latent variables or factors. For each pair of manifest variable $X_{i}$ and factor $F_{j}$, define
$$\tilde{l}_{i,j}^{2}:=\frac{l_{i,j}^{2}}{\text{Var}{\left(X_{i}\right)}}$$
Note that $\tilde{l}_{i,j}^{2}$ is the squared standardized loading of $X_{i}$ on $F_{j}$, which is also just equal to the squared correlation $\left(\text{Corr}\left(X_{i},F_{j}\right)\right)^{2}$. When $\text{Var}\left(X_{i}\right)=1$, such as when the manifest variables are standardized prior to factor analysis (which is often the case in practice), $\tilde{l}_{i,j}^{2}=l_{i,j}^{2}=\left(\text{Corr}\left(X_{i},F_{j}\right)\right)^{2}$.

Then, define $D_{f}:\left\{X_{1},X_{2},\dots,X_{M}\right\}\times\left\{X_{1},X_{2}\dots,X_{M}\right\}\to\left[0,1\right]$ as
\begin{equation}
D_{f}\left(X_{i},X_{j}\right):=1-\sqrt{\frac{1}{2}\sum_{k=1}^{T}{\left(\tilde{l}_{i,k}^{2}-\tilde{l}_{j,k}^{2}\right)^{2}}}
\end{equation}
Essentially, $D_{f}\left(X_{i},X_{j}\right)$ is the similarity between the two manifest variables in terms of their squared standardized loadings across all $T$ factors (i.e., squared correlations). We call this quantity the \textit{loading similarity}. Then, given a factor model $\mathscr{F}$, one can construct the \textit{loading similarity matrix} for $\mathscr{F}$, $\bm{\bm{\underline{U}}}=\left(D_{f}\left(X_{i},X_{j}\right)\right)_{1\leq i,j\leq M}$.

Note that $D_{f}$ measures how well the corresponding standardized loadings agree for two manifest variables $X_{i}$ and $X_{j}$. Two manifest variables are completely \textit{loading-wise similar} if and only if $D_{f}=1$ and completely \textit{loading-wise opposite} if and only if $D_{f}=0$, with larger values of $D_{f}$ indicating better agreement or higher degree of similarity. Clearly, $\bm{\bm{\underline{U}}}$ is symmetric and its diagonal entries are all equal to 1 (just like $\bm{\bm{\underline{Q}}}$). Instead of using the raw loadings $l_{i,j}$, the squared standardized loadings $\tilde{l}_{i,j}^{2}$ are used in order to control for the directions of the relationships (i.e., signs of the loadings) and the differing scales (i.e., levels of variability). A more detailed development of $D_{f}$ is shown in Appendix \hyperref[appendix:A]{A}.

\subsection{Interpretability Index}
At this point, we have described our criterion for interpretability and have mathematically defined the relevant similarity functions. Thus, we now introduce the \textit{interpretability index}, which operationalizes our criterion.

In a collection of factor models that are deemed equally plausible or statistically valid, those that are more interpretable are preferred as they are more practically useful and they convey ``real meaning" \parencite{semanticsFA}. Here, a factor model is considered more interpretable if the loadings and the semantics agree more. For a given model $\mathscr{F}$, define the multiset
$$Y:=\left\{\left(D_{s}\left(G\left(Q_{i}\right),G\left(Q_{j}\right)\right),D_{f}\left(X_{i},X_{j}\right)\right):\left(i,j\in\left\{1,2,\dots,M\right\}\right)\wedge\left(i<j\right)\right\}$$
An element of $Y$ is essentially an ordered pair containing the semantic similarity and the loading similarity between a pair of manifest variables. Then, define the interpretability index $\tau$ (or $\tau$-index) to be the Kendall Tau-$b$ correlation coefficient, from \textcite{stat132Ref}, mapped to the interval $\left[0,1\right]$.

That is,
\begin{equation}
\tau\left(Y\right):=\frac{N_{C}\left(Y\right)-N_{D}\left(Y\right)}{2\sqrt{N\left(Y\right)-N_{1}\left(Y\right)}\sqrt{N\left(Y\right)-N_{2}\left(Y\right)}}+\frac{1}{2},
\end{equation}
where $N_{C}\left(Y\right)$ is the number of concordant\footnote{A pair of ordered pairs $\left(x_{1},y_{1}\right)$ and $\left(x_{2}, y_{2}\right)$ is considered \textit{concordant} if $\left(x_{1}<x_{2}\right)\wedge\left(y_{1}<y_{2}\right)$ or $\left(x_{1}>x_{2}\right)\wedge\left(y_{1}>y_{2}\right)$, \textit{discordant} if $\left(x_{1}<x_{2}\right)\wedge\left(y_{1}>y_{2}\right)$ or $\left(x_{1}>x_{2}\right)\wedge\left(y_{1}<y_{2}\right)$, and neither concordant nor discordant otherwise.} pairs of elements in $Y$, $N_{D}\left(Y\right)$ is the number of discordant pairs of elements in $Y$, $N\left(Y\right)$ is the total number of pairs of elements in $Y$, $N_{1}\left(Y\right)$ is the total number of pairs of elements in $Y$ whose values at index $1$ are equal, and $N_{2}\left(Y\right)$ is the total number of pairs of elements in $Y$ whose values at index $2$ are equal.

\begin{figure}[H]
    \centering
    \includegraphics[width=0.5\textwidth]{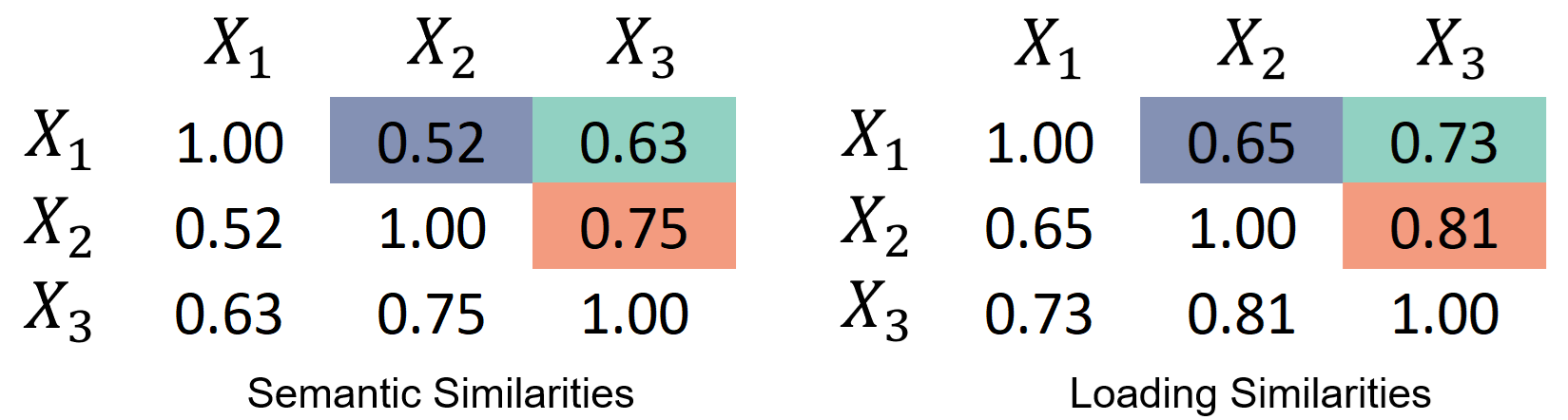}
    \caption{Semantic Similarity Matrix vs Loading Similarity Matrix}
    \label{fig:semantics-vs-loadings}
\end{figure}

The Kendall Tau-$b$ is mapped into the interval $\left[0,1\right]$ to make it easier to interpret. If $Y$ is strictly increasing (or at least, non-decreasing), then $\tau=1$, suggesting a strong agreement between semantics and loadings (the ideal case). For measuring agreement, monotonicity (e.g., using the Kendall Tau correlation) is considered instead of strict linearity (e.g., using the Pearson correlation) despite the similarity values themselves being ratio-level as it is not important to specify exactly the relationship between loading similarity and semantic similarity. Instead, we only require that they move in the same direction. Ideally, $D_{f}$ increases as $D_{s}$ increases. To understand this better, consider Figure \ref{fig:semantics-vs-loadings}.

Recall that from the two similarity functions, $D_{s}$ and $D_{f}$, we can construct two similarity matrices, $\bm{\underline{Q}}$ and $\bm{\bm{\underline{U}}}$. Both similarity functions are defined on any two pairs of variables $X_{i}$ and $X_{j}$ but $D_{s}$ is based on the semantics (i.e., meanings of the associated questions) while $D_{f}$ is based on the loadings. Essentially, the index is comparing these two similarity matrices and measuring their concordance by looking at the trend of the corresponding entries (which are color-coded in Figure \ref{fig:semantics-vs-loadings}).

There are various ways to measure the ``distance" between two matrices. For instance, we can use standard Euclidean distances (i.e., the Frobenius norm). However, we choose to leverage the Kendall Tau correlation element-wise since 1) we are interested in the trend among the similarity values but 2) we are not interested in the magnitudes of the similarity values themselves. It is unreasonable to expect that the semantic similarities exactly coincide with loading similarities (e.g., the semantic similarities may have different location and scale compared to loading similarities) --- it is also unnecessary for maximizing semantic coherence among manifest variables in a factor. Moreover, taking into account the magnitudes in $\tau$ would make the index depend\footnote{For example, suppose that $X_{1}$ and $X_{2}$ have corresponding questions that are semantically similar (compared to other pairs). The gap between the semantic similarity and the loading similarity would likely be larger if the communalities of $X_{1}$ and $X_{2}$ were different, even if their loadings were allocated in the same manner (e.g., high on $F_{1}$, low on $F_{2}$). An index that considers magnitudes would penalize this, even though we are interested only in the ``semantic coherence" (i.e., interpretability in terms of semantics).} on other characteristics of the model such as communalities and so on, which is not the goal of $\tau$. There are other goodness-of-fit indices for those purposes.

\begin{figure}[H]
  \centering
  \begin{subfigure}[c]{.45\textwidth}
      \centering
      \includegraphics[width=1\textwidth]{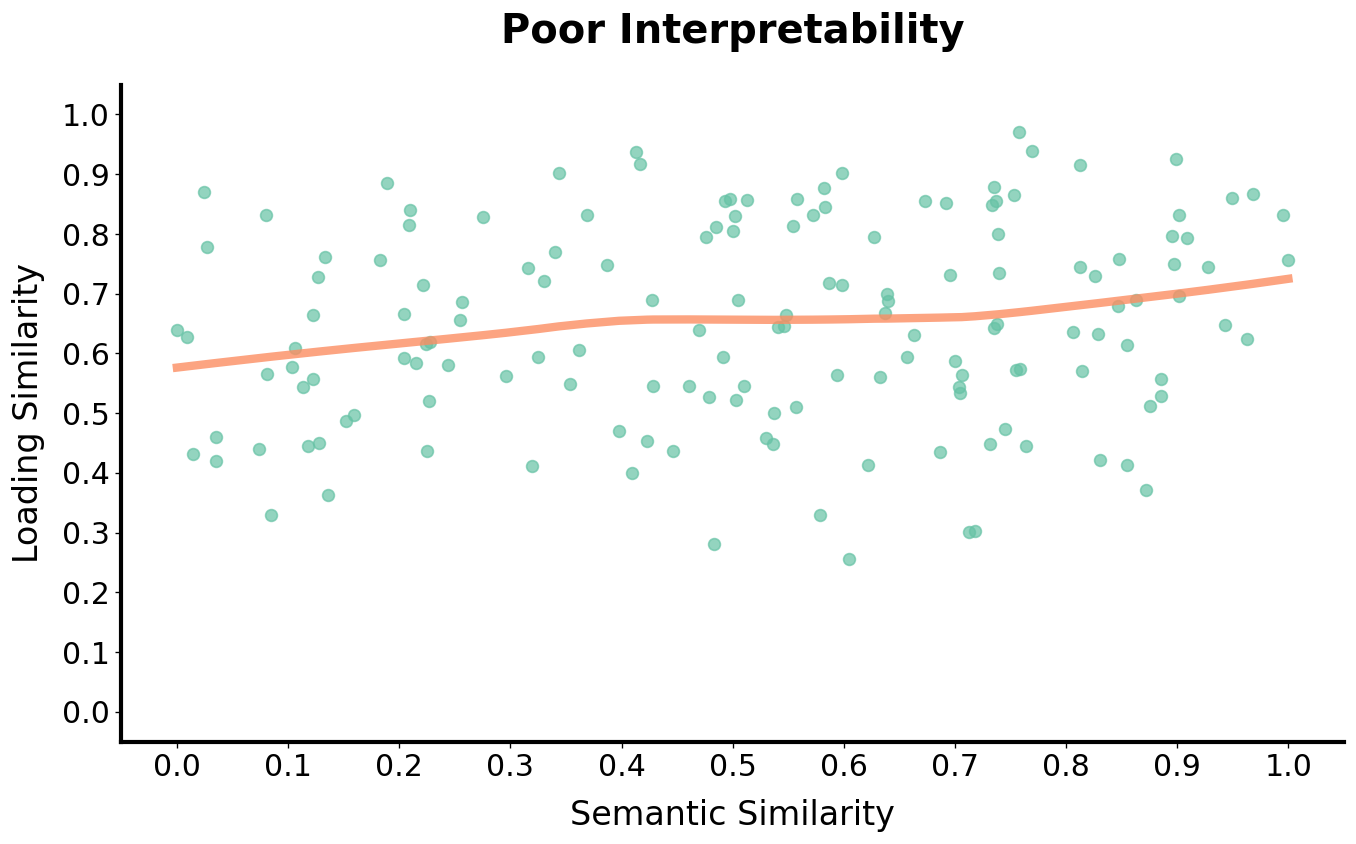}
      \caption{Poor Interpretability}
  \end{subfigure}%
  \begin{subfigure}[c]{.45\textwidth}
      \centering
      \includegraphics[width=1\textwidth]{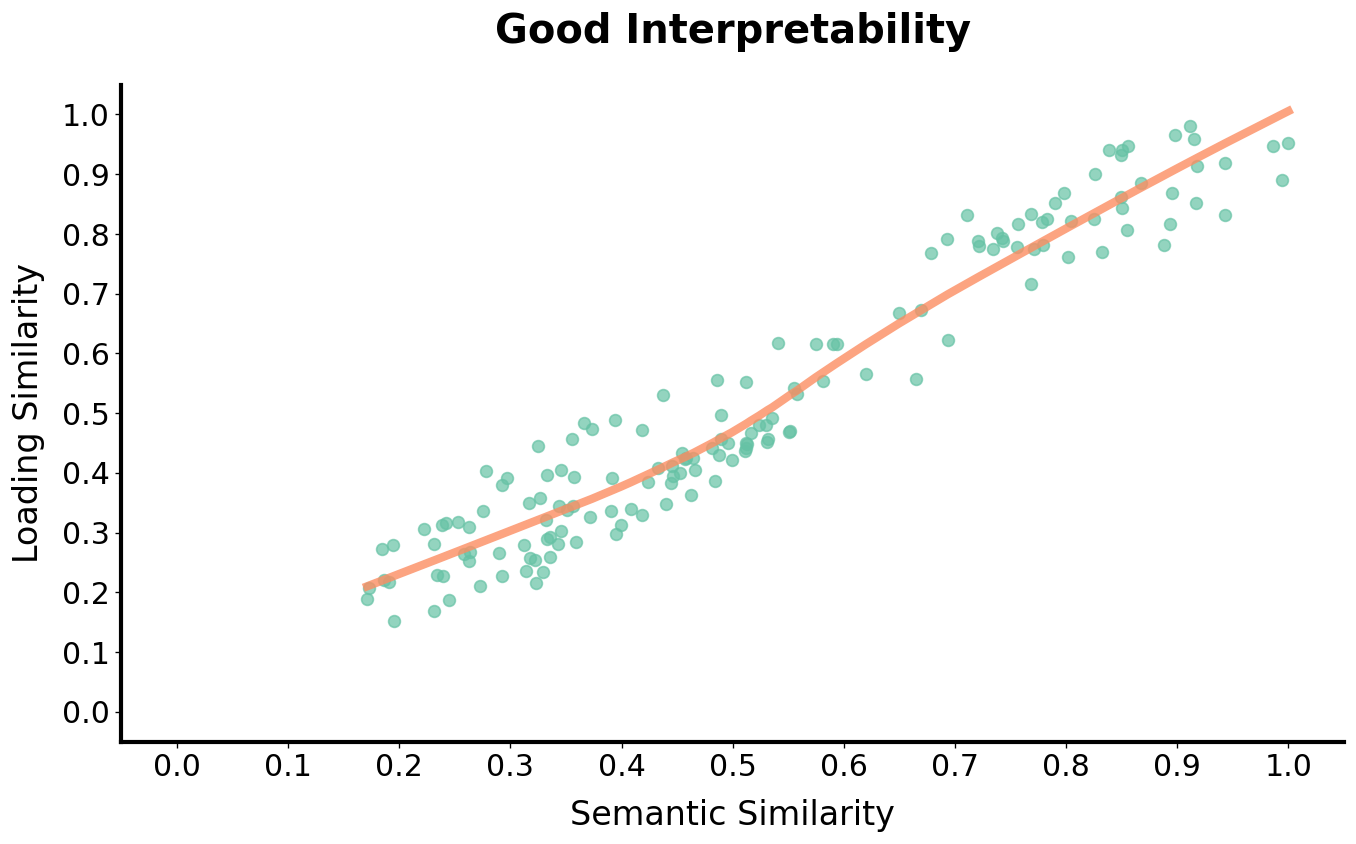}
      \caption{Good Interpretability\protect}
  \end{subfigure}
  \caption{The Interpretability Plot}
  \label{fig:poor-vs-good}
\end{figure}

\begin{figure}[H]
  \centering
  \begin{subfigure}[c]{.45\textwidth}
      \centering
      \includegraphics[width=1\textwidth]{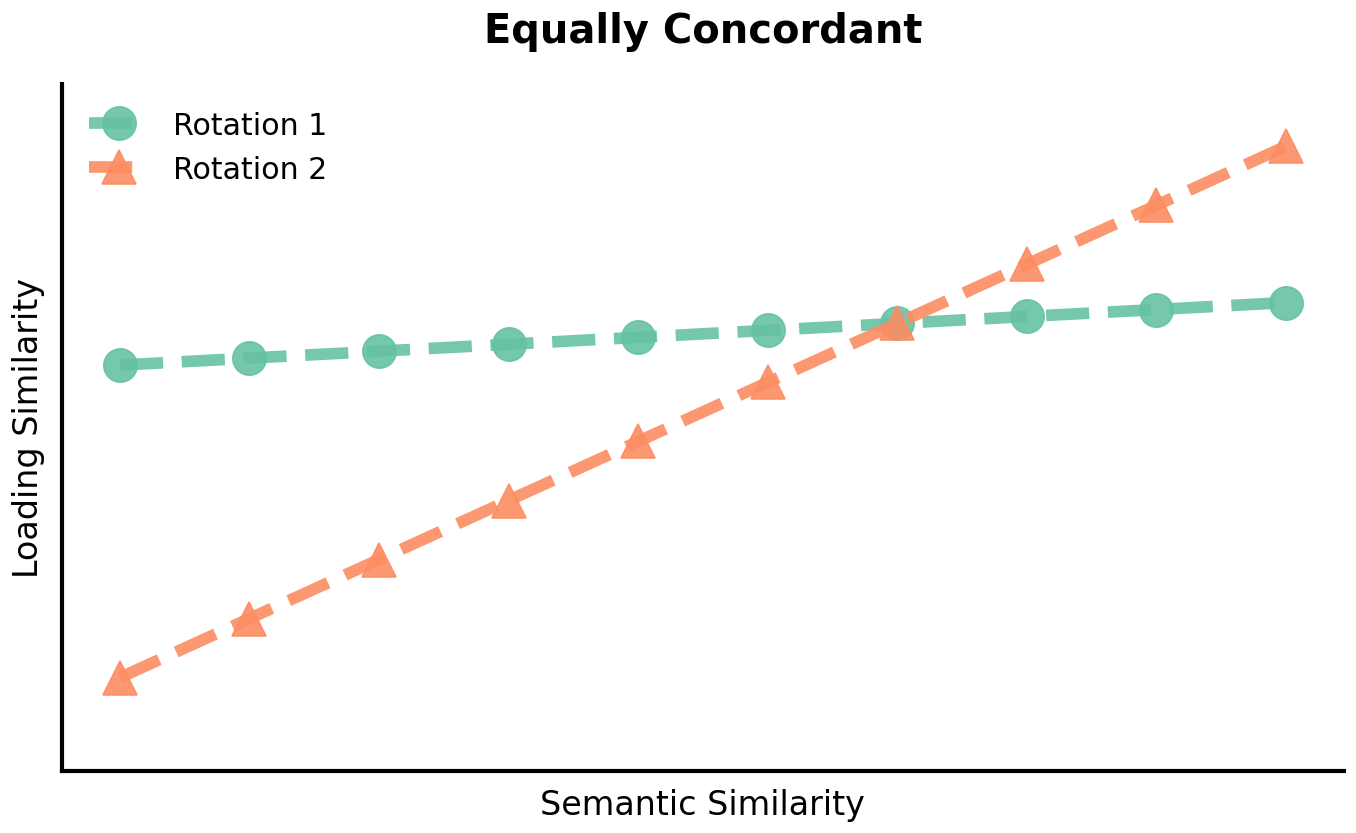}
      \caption{Equal Concordance}
  \end{subfigure}%
  \begin{subfigure}[c]{.45\textwidth}
      \centering
      \includegraphics[width=1\textwidth]{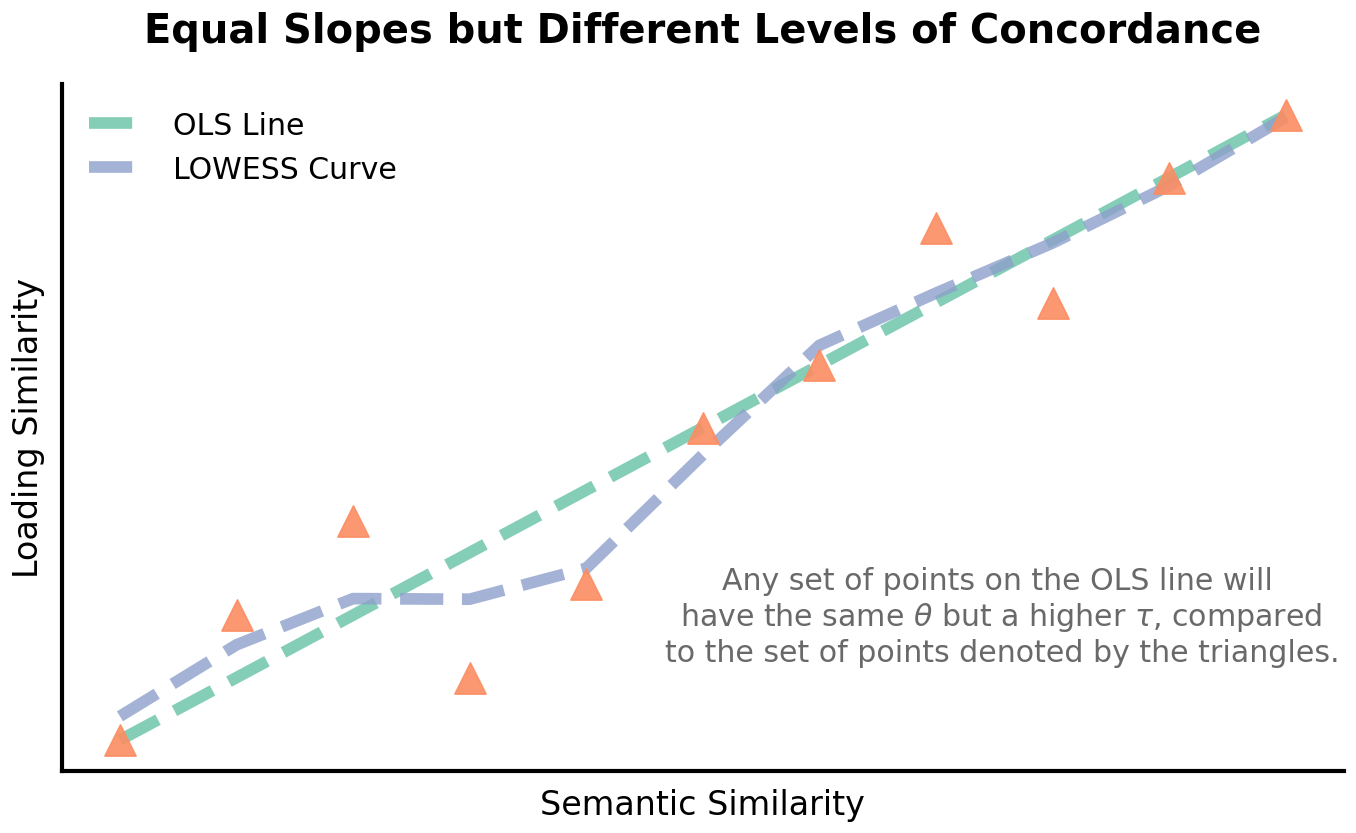}
      \caption{Equal Slopes}
  \end{subfigure}
  \caption{Modifying the $\omega$-index}
  \label{fig:improvement}
\end{figure}

To visualize the interpretability, one can construct a scatterplot of all $\left(x,y\right)\in Y$. To aid exploration further, one can then fit a locally weighted estimated scatterplot smoothing or LOWESS curve, from \textcite{lowess}, to see how loading similarity and semantic similarity vary together. We call this plot as the \textit{interpretability plot}. For instance, consider Figure \ref{fig:poor-vs-good}.
The plot on the left would indicate poor interpretability as there seems to be no relationship between semantic similarity and loading similarity. On the other hand, the plot on the right indicates good interpretability since there appears to be a positive monotone relationship. That is, as the semantic similarity increases, the loading similarity also increases. This would yield a high $\tau$ value. Although $\tau$ already operationalizes the interpretability criterion, we can still improve the index --- consider Figure \ref{fig:improvement}.

Particularly, consider first the left plot in the figure. It shows two sets of points, $Y_{1}$ (i.e., Rotation $1$) and $Y_{2}$ (i.e., Rotation $2$). Observe that $\tau\left(Y_{1}\right)=\tau\left(Y_{2}\right)=1$ and so, these two sets have the same $\tau$ value. Right now, the index suggests that they are equally interpretable. However, Rotation $2$ is arguably more interpretable since a unit change in semantic similarity corresponds to a larger change in loading similarity. This is a desirable characteristic (that can break ``ties" in $\tau$) because this suggests that groupings of manifest variables in terms of their semantics are more distinguishable or separable in terms of loadings (or loading patterns\footnote{Here, loading pattern refers to how the loadings are distributed across the factors for a given manifest variable. For example, $X_{1}$ may have a high loading on $F_{1}$, low loading on $F_{2}$, moderate loading on $F_{3}$, and so on.}). In other words, semantically similar manifest variables have highly similar loading patterns while semantically dissimilar ones have highly dissimilar patterns --- maximizing the slope would maximize the gaps among the ``clusters".

We can then augment the index by incorporating the slope based on the set $Y$. Particularly, we can utilize the \textit{ordinary least squares} (OLS) slope estimate on the set $Y$ as defined in \textcite{ols2013}, denoted by $\beta\left(Y\right)$, to measure how easily separable or distinguishable (in terms of loadings) semantically dissimilar manifest variables are. From this, note that greater values of $\left|\beta\left(Y\right)\right|$ indicate better ``separability" because they indicate larger changes in magnitude in loading similarity per unit change in semantic similarity.

However, aside from separability, recall that we also want a strong positive association between semantic similarity and loading similarity based on our premise for interpretability. Thus, instead of requiring just $\left|\beta\left(Y\right)\right|>0$, we require the stronger condition $\beta\left(Y\right)>0$. All things considered, larger values of $\beta\left(Y\right)$ indicate better interpretability. Note that the OLS slope estimate here is used \textit{purely for descriptive purposes} (e.g., no statistical inference). This means that Gauss-Markov assumptions or distributional (e.g., normality) assumptions are neither necessary nor relevant.

Now, observe that $\beta\left(Y\right)\in\left(-\infty,\infty\right)$ and is unbounded. To make it easier to use as (part of) an index, we can use the angle that corresponds to the slope (i.e., the angle that the line makes with the $x$-axis) and map it to $\left[0,1\right)$. Define $\theta\left(Y\right)$ as follows:
\begin{equation}
\theta\left(Y\right):=\frac{1}{\pi}\arctan{\left(\beta\left(Y\right)\right)}+\frac{1}{2}
\end{equation}
Since $\theta\left(Y\right)$ is just an order-preserving transformation $\beta\left(Y\right)$, higher values of $\theta$ also indicate better interpretability, for the same reasons described previously for $\beta\left(Y\right)$. Then, we can construct the final interpretability index, $\omega$ or the $\omega$-index, as the central tendency of $\tau\left(Y\right)$ and of $\theta\left(Y\right)$, particularly as the \textit{geometric mean}:
\begin{equation}
\omega:=\sqrt{\tau\left(Y\right)\theta\left(Y\right)}\in[0,1)
\end{equation}

\begin{figure}[H]
    \centering
    \includegraphics[width=0.55\textwidth]{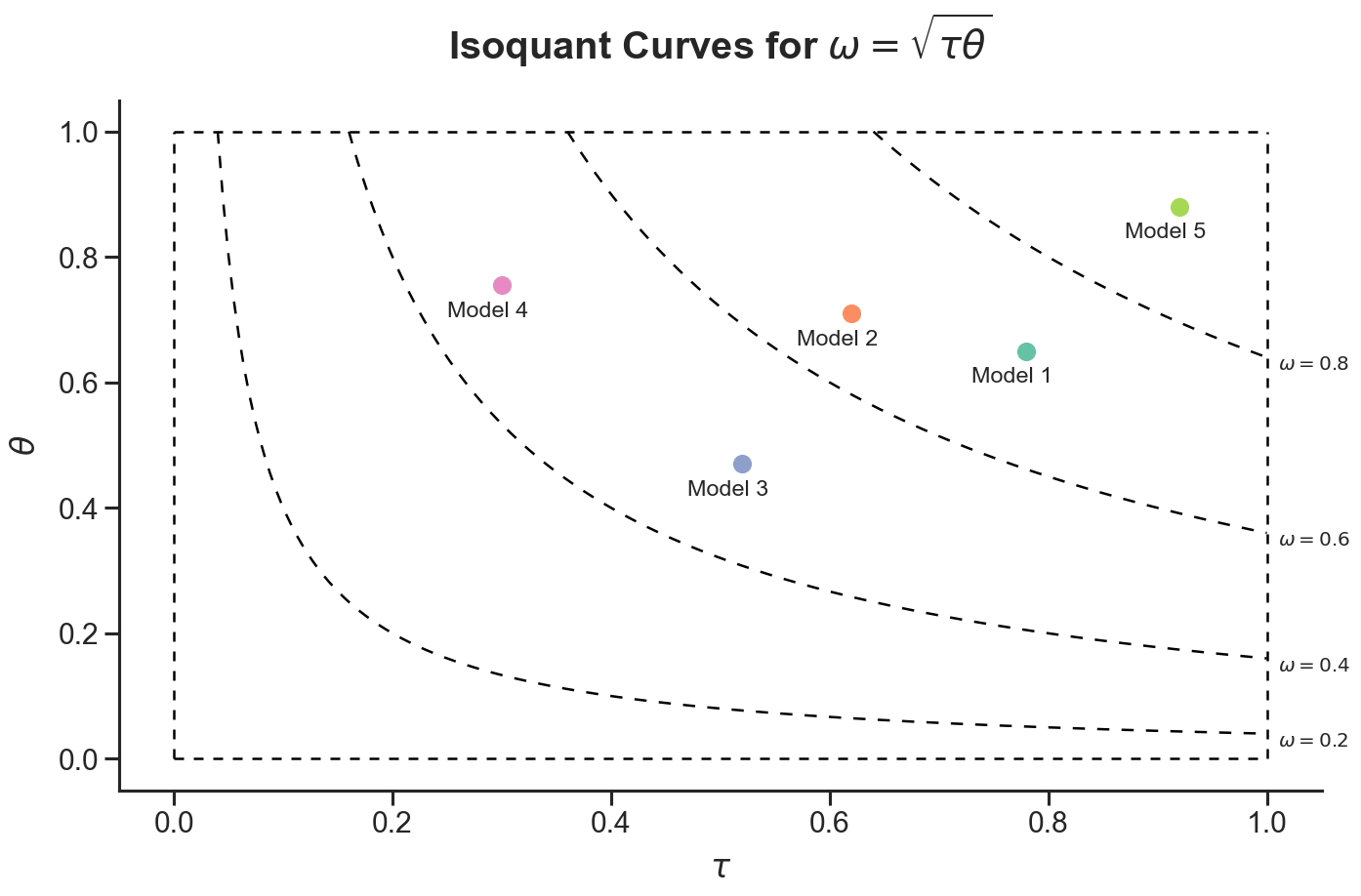}
    \caption{Isoquant Plot}
    \label{fig:isoquant}
\end{figure}

The geometric mean is used in order to incorporate the concept of \textit{diminishing marginal rate of substitution}. For example, if $\tau\left(Y\right)$ is already low, then one would would give up less of $\tau\left(Y\right)$ to increase $\theta\left(Y\right)$ since there will always be a point when $\tau\left(Y\right)$ just becomes ``too low'' no matter how high $\beta\left(Y\right)$ is (and vice versa). This formula, then, is the final form of $\omega$. Note that $\tau\left(Y\right)$ cannot be dropped from $\omega$ since $\theta\left(Y\right)$ can be the same for two sets where one is monotonically increasing (i.e., larger $\tau$, more closely meeting the interpretability criterion) and the other is not (i.e., smaller $\tau$), as suggested by the right plot in Figure \ref{fig:improvement}. Larger values of $\omega\in[0,1)$ indicate better interpretability. Ultimately, $\omega$ can be used for model selection, where we select the model or rotation that achieves the largest value for $\omega$. Although, the researcher can also easily look at the ``components" of $\omega$, $\tau$ and $\theta$, to assess and compare different models, as shown in the \textit{isoquant} plot in Figure \ref{fig:isoquant}.

It is important to note that $\omega$ is not necessarily a measure of absolute goodness-of-fit. It is entirely possible that the semantic similarities do not completely agree with the loading similarities (i.e., that $\omega$ is small) even in the absence of errors (e.g., model errors, errors in semantic similarities, etc.) or even in the population. The semantic similarity matrix is simply meant to be a form of \textit{a priori} information or ``targets'' (as in target rotation) that represent the researcher's expectation --- especially in the exploratory context. Instead, the index is generally more useful as a measure of relative goodness-of-fit (that is, for comparing two or more factor models that were fit using the same sample correlation matrix). With everything else equal, the model with the largest value for $\omega\in [0,1)$ is chosen.

Moreover, note that the index does not seek \textit{identical} loadings between a given pair of manifest variables, nor does it even seek to keep the loadings between any such pair \textit{similar}. Rather, the index \textit{compares similarities among pairs} --- it concerns itself with ``relative'' similarities instead of the absolute similarities. For example, it does not compare the loadings between $X_{1}$ and $X_{2}$ (i.e., absolute similarity). Instead, it compares the semantic similarities between $X_{1}$ and $X_{2}$, and $X_{1}$ and $X_{3}$ (i.e., relative similarity), then determines to which of $X_{2}$ and $X_{3}$ should $X_{1}$ be more loading-wise similar.
  
\subsection{Generalization}
As hinted previously, we can drop the assumption that the manifest variables have associated questions (i.e., that data is obtained by asking questions to respondents). Observe that the questions semantic similarity matrix $\bm{\underline{Q}}$ is essentially \textit{a priori} information about the model and the index measures how well the fitted factor model agrees with the prior information. In the case earlier, the prior information is the semantics of the questions.

However, $\bm{\underline{Q}}$ does not actually have to contain information about semantics (i.e., it need not be the semantic similarity matrix). It can be extended to a more general form $\bm{\underline{C}}=\left(c_{i,j}\right)_{1\leq i,j\leq M}$, which we call the \textit{prior matrix}. The prior matrix, then, provides a way to encode some \textit{a priori} information (i.e., the \textit{prior similarities}) and eventually, measure the concordance of the fitted factor model with the \textit{a priori} information. Unlike in the usual target rotation, the \textit{a priori} information here is encoded as pairwise similarities, not as specific values for loadings, which can be more intuitive when there are no specific assumptions on the exact values but there are weak expectations on which manifest variables should be alike (in terms of loading patterns). Conversely, a special example or case of the prior matrix is the semantic similarity matrix.

The investigator can construct the prior matrix to reflect how the manifest variables should be grouped together based on existing theory or domain knowledge (e.g., semantics) and then, use the $\omega$-index to quantify interpretability. To give an example, suppose that there are five ($5$) manifest variables --- $X_{1},X_{2},\dots,X_{5}$. Then, suppose that based on existing domain knowledge, $X_{1}$ and $X_{2}$ are expected to be grouped together while $X_{3}$, $X_{4}$, and $X_{5}$ are expected to constitute another group. A possible prior matrix $\bm{\underline{C}}$ would be
$$
\begin{bmatrix}
1 & 1 & 0 & 0 & 0 \\
1 & 1 & 0 & 0 & 0 \\
0 & 0 & 1 & 1 & 1 \\
0 & 0 & 1 & 1 & 1 \\
0 & 0 & 1 & 1 & 1
\end{bmatrix}
$$

Although, note that while $\bm{\underline{C}}$ expects $X_{1}$ and $X_{2}$ to be grouped together, it is not necessary that they are grouped together into exactly one factor. Just like in \textcite{enhancedInterpretability2020}, the $\omega$-index does not penalize cross-loadings (i.e., it does not seek simple structure or sparsity). It is possible that $X_{1}$ and $X_{2}$ have large enough loadings on multiple factors --- the only expectation is that they should have similar loading patterns (e.g., if $X_{1}$ loads high on $F_{1}$, then $X_{2}$ also loads relatively high on $F_{1}$), which may be the case in exploratory contexts.

In general, if variables $X_{i}$ and $X_{j}$ are preferred to be grouped together over variables $X_{i}$ and $X_{k}$, then set $c_{,j}>c_{i,k}$. If it is more important to group $X_{i}$ and $X_{j}$ together than it is to group $X_{m}$ and $X_{n}$ together, then set $c_{i,j}>c_{m,n}$. While the diagonal entries do not matter, $\bm{\underline{C}}$ must be a symmetric matrix. By convention, $\bm{\underline{C}}\in[0,1]^{M\times M}$. Essentially, $Y$ can be redefined as
$$Y:=\left\{\left(c_{i,j},D_{f}\left(X_{i},X_{j}\right)\right):\left(i,j\in\left\{1,2,\dots,M\right\}\right)\wedge\left(i<j\right)\right\}$$
Moreover, we refer to pairs with large $c_{i,j}$ values as \textit{prior-similar} and pairs with small $c_{i,j}$ values as \textit{prior-dissimilar}. Naturally, $c_{i,j}>c_{i,k}$ means that $X_{i}$ is more prior-similar to $X_{j}$ than it is to $X_{k}$.

Now, as another example, consider the matrix
$$
\begin{bmatrix}
1 & 1 & 0 & 0 & 0 \\
1 & 1 & 0 & 0 & 0 \\
0 & 0 & 1 & 1 & 0.5 \\
0 & 0 & 1 & 1 & 0.5 \\
0 & 0 & 0.5 & 0.5 & 1
\end{bmatrix}
$$
This prior matrix suggests that $X_{1}$ and $X_{2}$ should be ``grouped" together and that $X_{3}$, $X_{4}$, and $X_{5}$ should constitute another group. However, among $X_{3}$, $X_{4}$, and $X_{5}$, it is expected that $X_{5}$ will not ``belong as much" to the group compared to the other two. In other words, while $X_{5}$ is grouped together with $X_{3}$ and $X_{4}$, it is only relatively weakly similar (compared to how similar $X_{3}$ and $X_{4}$ are).

Another way to extend the prior matrix $\bm{\underline{C}}$ is to modify it to allow partial specifications. To do so, we can simply redefine $Y$ as 
$$Y:=\left\{\left(c_{i,j},D_{f}\left(X_{i},X_{j}\right)\right):\left(i,j\in\left\{1,2,\dots,M\right\}\right)\wedge\left(i<j\right)\wedge\left(c_{i,j}\neq \Delta\right)\right\}$$
Then, $\omega$ is computed ignoring pairs of manifest variables where $c_{i,j}=\Delta$. For example, suppose that there are six (6) variables $X_{1},\dots,X_{6}$. Next, suppose that $X_{1}$ and $X_{3}$ should be grouped together and $X_{4}$ and $X_{6}$ should be grouped together. However, it is unknown how $X_{2}$ and $X_{5}$ should be grouped (i.e., there is no prior information about these two variables). Then, a possible prior matrix $\bm{\underline{C}}$ would be
$$
\begin{bmatrix}
1 & \Delta & 1 & 0 & \Delta & 0 \\
\Delta & \Delta & \Delta & \Delta & \Delta & \Delta \\
1 & \Delta & 1 & 0 & \Delta & 0 \\
0 & \Delta & 0 & 1 & \Delta & 1 \\
\Delta & \Delta & \Delta & \Delta & \Delta & \Delta \\
0 & \Delta & 0 & 1 & \Delta & 1 
\end{bmatrix}
$$
The index $\omega$ still provides a measure of concordance with the prior information or expectation represented in $\bm{\underline{C}}$ as it still acts on the available information in a way similar to how it does for a fully specified $\bm{\underline{C}}$. Although this time, it is invariant with respect to models whose only differences are in variable relationships with no prior information (i.e., relationships whose prior information is $\Delta$).

One can also use something based on the semantic similarity matrix $\bm{\underline{Q}}$, but not $\bm{\underline{Q}}$ itself directly. For instance, since $\bm{\underline{Q}}$ is technically a similarity matrix, one can cluster the associated questions (and by extension, the manifest variables) based on this similarity matrix. This can be done using general clustering methods, such as affinity propagation from \textcite{affinityProp} and $k$-medoids from \textcite{kmedoids}. Then, one can define block matrices (e.g., one block corresponds to one cluster) to construct the final prior matrix $\bm{\underline{C}}$ based on the obtained clusters.

Ultimately, observe that the factor model is estimated by using the correlation matrix, which can be thought of as a similarity matrix. The prior matrix $\bm{\underline{C}}$ (such as the semantic similarity matrix) is simply another similarity matrix that is used to resolve the rotational indeterminacy. Notably, some specifications of $\bm{\underline{C}}$ can be considered ``equivalent". For example, $\omega$ does not change when applying the same translation to each element of $\bm{\underline{C}}$ because $\tau$ and $\theta$ are invariant to translations. On a similar note, the loading matrix that maximizes $\omega$ does not change when applying the same order-preserving affine transformation to each element of $\bm{\underline{C}}$. In the case where $\bm{\underline{C}}$ is constructed by computing semantic similarities, this means that the specific choice for the embedder is inconsequential if the set of embedders considered provide semantic similarities that are (approximately) identical up to translations or order-preserving affine transformations.

\subsection{Rotation}
Because $\omega$ is an index or criterion, a rotation method can be built based on it --- one that maximizes $\omega$. This allows for a more ``active" way of incorporating prior information (i.e., directly influencing the loading matrix instead of simply computing it for an existing loading matrix). This can be implemented as an optimization problem. Recall that
$$
\omega=\sqrt{\tau\left(Y\right)\theta\left(Y\right)},
$$
where
$$Y=\left\{\left(c_{i,j},D_{f}\left(X_{i},X_{j}\right)\right):\left(i,j\in\left\{1,2,\dots,M\right\}\right)\wedge\left(i<j\right)\right\}$$
Observe that $\omega$ can be expressed as a function of the rotation matrix, which we assume to be orthogonal, when the unrotated loadings $\bm{\underline{L}}$ and the prior matrix $\bm{\underline{C}}$ are known: $\omega=f\left(\bm{\underline{R}};\bm{\underline{L}},\bm{\underline{C}}\right)$. We call the rotation that maximizes $\omega$ as pairwise target rotation or the \textit{priorimax rotation}, which is the following optimization problem:
\begin{maxi*}|s|
{\bm{\underline{R}}}
{\omega}
{}
{}
\addConstraint{\bm{\underline{R}}\bm{\underline{R}}^{\prime}=\bm{\underline{I}}}
\end{maxi*}
It is not necessary that the rotation matrix is orthogonal for maximizing $\omega$, as $\omega$ is applicable even for oblique models. However, as a starting point, we consider only orthogonal rotations in this paper.

The optimization problem can be solved using the Constrained Optimization By Quadratic Approximations (COBYQA) algorithm, a derivative-free optimization algorithm, from \textcite{cobyqa}, as implemented in \codify{scipy} by \textcite{scipy}. COBYQA is generally fast (compared to global optimization algorithms) but it is technically designed for local optimization and so, we use multiple random starts to increase the chances of reaching the global maximum (or near the global maximum). Some reduction in precision is accepted in favor of speed, as the rotation method is primarily exploratory in nature, where computational efficiency is prioritized over exactness.

\subsection{Theoretical Characterization}
Earlier, we noted that the priorimax rotation does not necessarily seek simple structure. To see why more clearly, we compare the priorimax rotation to the orthomax family of rotations characterized in \textcite{jennrich2003}. In particular, we directly contrast our approach to the \textit{varimax} rotation from \textcite{Kaiser1958}.

Recall that the orthomax criterion (under othorgonal rotation), as discussed in \textcite{Crawford1970} and in \textcite{jennrich2003}, is given by
$$
O_{\gamma}:=M\sum_{i=1}^{T}{\sum_{j=1}^{M}{l_{j,i}^{4}}}-\gamma\sum_{i=1}^{T}{\left(\sum_{j=1}^{M}{l_{j,i}^{2}}\right)^{2}}
$$
As noted in \textcite{Crawford1970}, a reasonable range for $\gamma$ is $\left[0,M\right]$, although $\gamma$ can technically be any real value. In particular, $\gamma=1$ retrieves the (raw) varimax criterion from \textcite{Kaiser1958}:
$$
O_{1}:=M\sum_{i=1}^{T}{\sum_{j=1}^{M}{l_{j,i}^{4}}}-\sum_{i=1}^{T}{\left(\sum_{j=1}^{M}{l_{j,i}^{2}}\right)^{2}}=M\sum_{i=1}^{T}\left(\sum_{j=1}^{M}{l_{j,i}^{4}}-\left(\sum_{j=1}^{M}{l_{j,i}^{2}}\right)^{2}\right)
$$
In Kaiser's original formulation, the raw varimax criterion is given by
$$
V^{*}:=\frac{O_{1}}{M^{2}}=\sum_{i=1}^{T}\left(\frac{1}{M}\sum_{j=1}^{M}{l_{j,i}^{4}}-\left(\frac{1}{M}\sum_{j=1}^{M}{l_{j,i}^{2}}\right)^{2}\right)
$$

In any case, $O_{1}$ or equivalently, $V^{*}$, is maximized to hopefully orient the loading matrix to a simple structure \parencite{Crawford1970}. As described by \textcite{Kaiser1958}, $V^{*}$ is simply the sum of the variances of the squared loadings of each factor. Moreover, it is well-known that the variance can also be expressed as a sum of the squared pairwise differences, as described in \textcite{heffernan}, so that
$$
V^{*}=\sum_{i=1}^{T}\text{Var}_{j}\left(l_{j,i}^{2}\right)=\frac{1}{M^{2}}\sum_{j=1}^{M-1}\sum_{k=j+1}^{M}\sum_{i=1}^{T}\left(l^{2}_{j,i}-l_{k,i}^{2}\right)^{2}
$$
Now, note that
$$
\sum_{k=1}^{T}\left(\tilde{l}_{i,k}^{2}-\tilde{l}_{j,k}^{2}\right)^{2}=2\left(1-D_{f}\left(X_{i},X_{j}\right)\right)^{2}
$$
for any valid $i$ and $j$. Hence, when the manifest variables are standardized (or when all manifest variables have variances equal to $1$), which is almost always the case in practice, we can express $V^{*}$ as
$$
V^{*}=\frac{2}{M^{2}}\sum_{i=1}^{M-1}\sum_{j=i+1}^{M}\left(1-D_{f}\left(X_{i},X_{j}\right)\right)^{2}
$$
In contrast, maximizing the $\theta$ component of the objective function $\omega$ of the priorimax rotation is equivalent to maximizing
$$
\beta^{*}:=\frac{M\left(M-1\right)}{2}\sum_{i=1}^{M-1}\sum_{j=i+1}^{M}\left(c_{i,j}D_{f}\left(X_{i},X_{j}\right)\right)-\left(\sum_{i=1}^{M-1}\sum_{j=i+1}^{M}c_{i,j}\right)\left(\sum_{i=1}^{M-1}\sum_{j=i+1}^{M}D_{f}\left(X_{i},X_{j}\right)\right)
$$
Let $\overline{c}:=\frac{2}{M(M-1)}\sum_{i<j}{c_{i,j}}$. That is, $\overline{c}$ is the average of all prior similarities $c_{i,j}$. Then, it is easy to see that
$$
\beta^{*}=\frac{M\left(M-1\right)}{2}\sum_{i=1}^{M-1}\sum_{j=i+1}^{M}\left(\left(c_{i,j}-\overline{c}\right)D_{f}\left(X_{i},X_{j}\right)\right)
$$
Thus, varimax maximizes
\begin{equation}
V^{\star}\propto\sum_{i=1}^{M-1}\sum_{j=i+1}^{M}\left(1-D_{f}\left(X_{i},X_{j}\right)\right)^{2}
\end{equation}
while priorimax maximizes
\begin{equation}
\beta^{*}\propto\sum_{i=1}^{M-1}\sum_{j=i+1}^{M}\left(\left(c_{i,j}-\overline{c}\right)D_{f}\left(X_{i},X_{j}\right)\right)
\end{equation}

Because 
$D_{f}\left(X_{i},X_{j}\right)\in\left[0,1\right]$
for all valid 
$i$ and $j$,
varimax essentially aims to minimize the loading similarity across pairs of manifest variables with equal priority, making the best effort in orienting the loading matrix towards a (perfect) simple structure. Indeed, if an orthogonal rotation of the initial loading matrix follows perfect simple structure, then varimax (or any orthomax rotation where $0\leq \gamma \leq 1$, for that matter) recovers the structure up to permutation and sign changes \parencite{jennrich2003}.

On the other hand, priorimax aims to make some pairs of manifest variables loading-wise similar (e.g., those that are semantically similar) and other pairs loading-wise dissimilar (e.g., those that are semantically dissimilar). More precisely, it actively maximizes the loading similarities of pairs with $\left(c_{i,j}-\overline{c}\right)>0$ (i.e., pairs that are more prior-similar than average) and actively minimizes for pairs with $\left(c_{i,j}-\overline{c}\right)<0$. Then, among prior-similar (prior-dissimilar) pairs, priorimax focuses more on maximizing (minimizing) the loading similarities of pairs with large values for $\left|c_{i,j}-\overline{c}\right|$ than maximizing (minimizing) those with small absolute coefficients.

Hence, unlike varimax that minimizes \textit{all} loading similarities, priorimax minimizes \textit{only} the loading similarities of prior-dissimilar pairs and maximizes the rest, as made clear by $\left(7\right)$ and $\left(8\right)$. Ultimately, one can expect that priorimax may be better suited when true cross-loadings are present and prior similarities are available, and less appropriate when (perfect) simple structure truly exists. Later on, we examine this via a Monte Carlo experiment.

Finally, note that we did not consider the $\tau$ component of $\omega$ from the priorimax rotation in the comparison. However, $\tau$ should largely be comonotonic (or at least, not conflicting) with $\beta$, so that zeroing in on $\theta$ simplifies the analysis while preserving the essence of priorimax. Arguably, one can view $\theta$ as the ``high-level" driver of priorimax that describes the overall trend between prior and loading similarities, while $\tau$ as the ``low-level" driver that describes the point-by-point trend. While an ablation study of $\omega$ is beyond the current scope, evaluating $\tau$ and $\theta$ separately could provide useful insights. In fact, we suggest examining variations of $\omega$, possibly one that depends only on $\beta$ (i.e., matching the simplified case in this section), and briefly discuss such in Section \ref{sec:5}.

\subsection{Software Implementation}
For the software implementation of the proposed index, rotation, and visualizations, we developed the \codify{interpretablefa} package, which extends \codify{FactorAnalyzer} class from the \codify{Python} package \codify{factor\_analyzer}. The package \codify{interpretablefa} and its documentation are available on the Python Package Index, \url{https://pypi.org/project/interpretablefa/}.

\section{Simulation and Empirical Results}\label{sec:4}
\subsection{Simulation Design}
In order to characterize some properties of $\omega$ and the (orthogonal) priorimax rotation, we conducted a Monte Carlo experiment, which primarily considered three experimental factors: the type of loading matrix, the magnitude of the specific variances, and the level of perturbation in the prior matrix. For every experimental condition (i.e., a design point), we simulated $1000$ replicates. The possible values (i.e., levels) for the experimental factors are:
\begin{itemize}
    \item Loading matrix: Simple, Generic
    \item Specific variance: Small, Medium, Large
    \item Perturbation level in the prior matrix: $0.0,~0.25,~0.5~,0.75,~1.0$
\end{itemize}
Similar to the simulation study from \textcite{enhancedInterpretability2020}, we considered a specific case here --- models with $18$ manifest variables and $3$ factors, with a sample size of $300$. Each manifest variable had a mean of $0$ (i.e., $\mathbb{E}\left(X_{i}\right)=0$ for all $i$) and a total variance of $1$ (i.e., $\text{Var}\left(X_{i}\right)=1$ for all $i$). The specifics of the design (e.g., what ``small" or ``simple" mean) are discussed over the next few paragraphs.

Now, we considered two kinds of loading matrices: simple and generic. In the case of simple loading matrices, the loading matrix followed a simple structure such that each manifest variable loaded high only one factor and low on the rest (i.e., the loading matrix was sparse). Moreover, each factor had an equal number of high-loading manifest variables (i.e., $6$ each). In other words, the number of manifest variables ``assigned" to each factor is the same. On the other hand, generic loading matrices were not constrained to follow a simple structure --- it allowed for cross-loadings. Manifest variables did not necessarily have the same communalities but the specific variance of each variable was varied from small to large. The exact algorithm used to generate loading matrices are Algorithms \ref{alg:setting-psis}, \ref{alg:generate-simple-loadings}, and \ref{alg:generate-generic-loadings}, which are shown in Appendix \ref{appendix:B}. However, at a high-level, the procedure is as follows:
\begin{enumerate}
    \item Decide on the general level of the specific variances (i.e., small, medium, or large) and on the kind of loading matrix (i.e., simple or generic).
    \item If the level of the specific variances is small, draw the values for the specific variances from $U\left(0.1,0.3\right)$. For medium and large specific variances, the random draws are taken from $U\left(0.3,0.5\right)$ and $U\left(0.5,0.7\right)$, respectively. Note that the general level (e.g., small) is the same for all manifest variables but the exact specific variances differ.
    \item To generate a simple loading matrix:
    \begin{enumerate}
        \item Assign each manifest variable to exactly one factor, such that each factor has the same number of manifest variables assigned to it. Note that assignment can be done arbitrarily without loss of generality.
        \item For every loading $l_{i,j}$, if the manifest variable $X_{i}$ is assigned to factor $F_{j}$, the size of the loading is large. Otherwise, the size is medium with $25\%$ chance and small with $75\%$ chance.
        \item For every loading, if the size is large, assign to it a value taken from $U\left(0.8,1.0\right)$. Otherwise, the value is drawn from $U\left(0.4,0.6\right)$ for medium-sized loadings and from $U\left(0,0.2\right)$ from small-sized loadings.
        \item For every loading, decide on its sign by randomly drawing from $\left\{+,-\right\}$ with equal probabilities.
        \item Considering the specific variance for the corresponding manifest variable, normalize each row of the loading matrix such that the total variance of the manifest variable is $1$.
        \item The constructed matrix is a simulated loading matrix.
    \end{enumerate}
    \item To generate a generic loading matrix:
    \begin{enumerate}
        \item For every loading, assign a size --- large, medium, or small --- with equal probabilities.
        \item Follow steps $3\left(\text{c}\right)$ to $3\left(\text{f}\right)$.
    \end{enumerate}
\end{enumerate}

Then, using the fact that $\bm{\underline{X}}=\bm{\underline{L}}\bm{\underline{F}}+\bm{\underline{\varepsilon}}$, sample data were generated using Algorithm \ref{alg:generate-data}, shown in Appendix \ref{appendix:C}. In summary, to generate the data matrix $\bm{\underline{D}}\in\mathbb{R}^{N\times M}$, where each row is an observation and each column represents values for a manifest variable:
\begin{enumerate}
    \item For every observation $n$:
    \begin{enumerate}
        \item Generate realized values for the (orthogonal) factors by drawing a random sample $f_{1},f_{2},\dots,f_{T}$ from $\mathcal{N}\left(0,1\right)$. Note that because the factors are orthogonal, $\text{Cov}\left(F_{i},F_{j}\right)=0$ for any $i<j$ and $i,j\in\left\{1,2,\dots,T\right\}$. Thus, the factor scores are drawn independently.
        \item For every manifest variable $X_{i}$:
        \begin{enumerate}
            \item Draw $\varepsilon_{n.i}$ from $\mathcal{N}\left(0,1\right)$ for the error term (i.e., the contribution of the specific factor), where $\psi_{i}$ is the specific variance of $X_{i}$.
            \item The value of the observation $X_{i}$ is given by $x_{n,i}:=\sum_{j=1}^{T}{l_{i,j}f_{j}}+\varepsilon_{n,i}$, where $l_{i,j}$ is the loading of $X_{i}$ on $F_{j}$.
        \end{enumerate}
        \item The vector $\begin{bmatrix}x_{n,1} & x_{n,2} & \dots & x_{n,M}\end{bmatrix}$ is the $n$th observation.
    \end{enumerate}
    \item The matrix $\left(x_{i,j}\right)_{1\leq i \leq N,1\leq j \leq M}$ is the data matrix.
\end{enumerate}

For simulating a prior matrix, we considered perturbed versions of the true loading similarity matrix produced by computing pairwise loading similarities with $D_{f}$, as defined before. Perturbations represented noise in the prior matrix and the level of perturbation was varied.  Absence of perturbation represented the ideal case where the prior similarities (e.g., semantic similarities) served as perfect proxies for the loadings, which is unrealistically favorable for priorimax yet provides a baseline for comparison. Meanwhile, larger levels of perturbation represented more realistic scenarios where noise and inaccuracies were present in the prior similarities (e.g., when estimating semantic similarities, when prior similarities do not exactly match loading similarities, and so on). The exact algorithm is Algorithm \ref{alg:generating-prior}, shown in Appendix \ref{appendix:D}. However, in general, to generate a prior matrix:
\begin{enumerate}
    \item Decide on the level of perturbation, $\delta\in\left[0,1\right]$.
    \item Construct the loading similarity matrix $\bm{\underline{U}}$ based on the true (simulated) loading matrix.
    \item Perturb the loading similarity matrix by adding to each value in the upper triangle a random draw from $U\left(-\delta,\delta\right)$ and then, copying each value in the upper triangle to the corresponding element in the lower triangle to maintain symmetry. The draw is done per element or value.
    \item Shift the perturbed matrix (i.e., add a constant to every value) such that the smallest value in the matrix becomes $0$.
    \item Scale the shifted perturbed matrix (i.e., divide each value by a constant) such that the largest value is $1$.
    \item Set all diagonal values of the shifted and scaled perturbed matrix to $1$. This matrix is the (simulated) prior matrix.
\end{enumerate}

As for the method used for estimating the correlation matrix (by extension, the loading matrix), MinRes from \textcite{Harman1966} was utilized. In addition to varying the communalities and the levels of perturbation, several orthogonal rotations were also considered (i.e., varimax, equamax, and quartimax), aside from the unrotated loading matrix. Moreover, the primary metrics monitored in the simulation study were the $\omega$-index and the root mean square error (RMSE):
$$
\text{RMSE}\left(\hat{\bm{\underline{L}}}_{M\times T},\bm{\underline{L}}_{M\times T}\right):=\sqrt{\frac{1}{MT}\sum_{i=1}^{M}\sum_{j=1}^{T}\left(\hat{l}_{i,j}-l_{i,j}\right)^{2}},
$$
where $\hat{\bm{\underline{L}}}_{M\times T}$ is the estimated loading matrix and $\bm{\underline{L}}_{M \times T}$ is the true loading matrix. Note that as a consequence of rotational indeterminacy, factor models also have permutation indeterminacy and sign indeterminacy. These two types of indeterminacies do not materially change the interpretation of each factor (or the process of interpretation) at all and so to control for these, we used an adjusted version of the RMSE:
$$
\text{RMSE}_{\text{Adj.}}\left(\hat{\bm{\underline{L}}}_{M\times T},\bm{\underline{L}}_{M\times T}\right):=\min_{\bm{\underline{P}}_{T\times T},\bm{\underline{S}}_{T\times T}}{\left(\text{RMSE}\left(\hat{\bm{\underline{L}}}\bm{\underline{P}},\bm{\underline{L}}\bm{\underline{S}}\right)\right)},
$$
where $\bm{\underline{P}}$ is a permutation matrix and $\bm{\underline{S}}$ is a signature matrix. Finally, observe that $\text{RMSE}_{\text{Adj.}}$ naturally becomes smaller as specific variances become larger, since the communalities become smaller. Hence, to make comparisons of RMSE across different levels of communalities more appropriate, we scaled $\text{RMSE}_{\text{Adj.}}$:
$$
\text{RMSE}_{\text{Scl.Adj.}}\left(\hat{\bm{\underline{L}}}_{M\times T},\bm{\underline{L}}_{M\times T}\right):=\frac{\text{RMSE}_{\text{Adj.}}\left(\hat{\bm{\underline{L}}}_{M\times T},\bm{\underline{L}}_{M\times T}\right)}{\sqrt{\frac{2}{T}\left(1-\overline{\psi}\right)}},
$$
where $\overline{\psi}$ is the (common) expected value of the specific variances. That is, $\overline{\psi}=0.2$, $\overline{\psi}=0.4$, and $\overline{\psi}=0.6$ when the specific variance level is ``small", ``medium", and ``large", respectively. The denominator is the $\text{RMSE}_{\text{Adj.}}$ value when the true loading matrix and the estimated loading matrix are ``completely non-overlapping", assuming that $\psi_{1}=\psi_{2}=\dots=\psi_{M}=\overline{\psi}$ for both the true and the estimated models. That is, for every pair of corresponding rows from $\bm{\hat{\underline{L}}}$ and $\bm{\underline{L}}$, there is no column (or position) where both rows have a non-zero value, even after any sign or column permutation.

Generally, one replicate consisted of simulating the loading matrix, constructing the data matrix based on the loading matrix, and simulating the prior matrix, taking into account the experimental condition. Then, in a replicate, various orthogonal rotations were considered (including the orthogonal priorimax rotation and the identity rotation) and the primary metric values were noted. To summarize the $1000$ replicates of each given experimental condition, the mean, first quartile, and third quartiles (which correspond to the main series lines, lower bands, and upper bands, respectively, in the visualizations) of the $\omega$-index and the scaled adjusted RMSE across the replicates were taken. The experiment was implemented in \codify{Python 3} with \codify{interpretablefa}.

\subsection{Simulation Results}
\begin{figure}[H]
  \centering
  \begin{subfigure}[c]{.5\textwidth}
      \centering
      \includegraphics[width=1\textwidth]{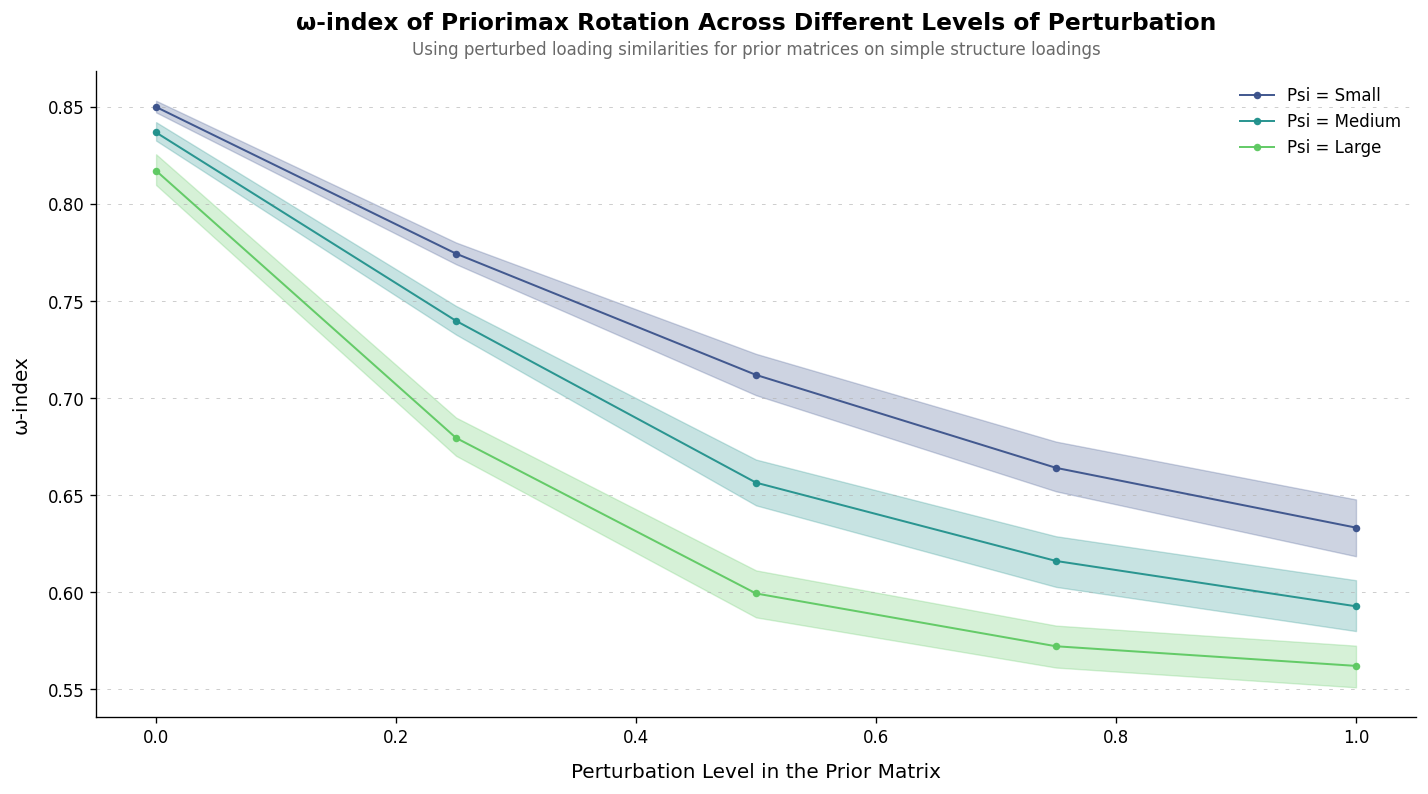}
      \caption{$\omega$ on Simple Loading Matrices}
  \end{subfigure}%
  \begin{subfigure}[c]{.5\textwidth}
      \centering
      \includegraphics[width=1\textwidth]{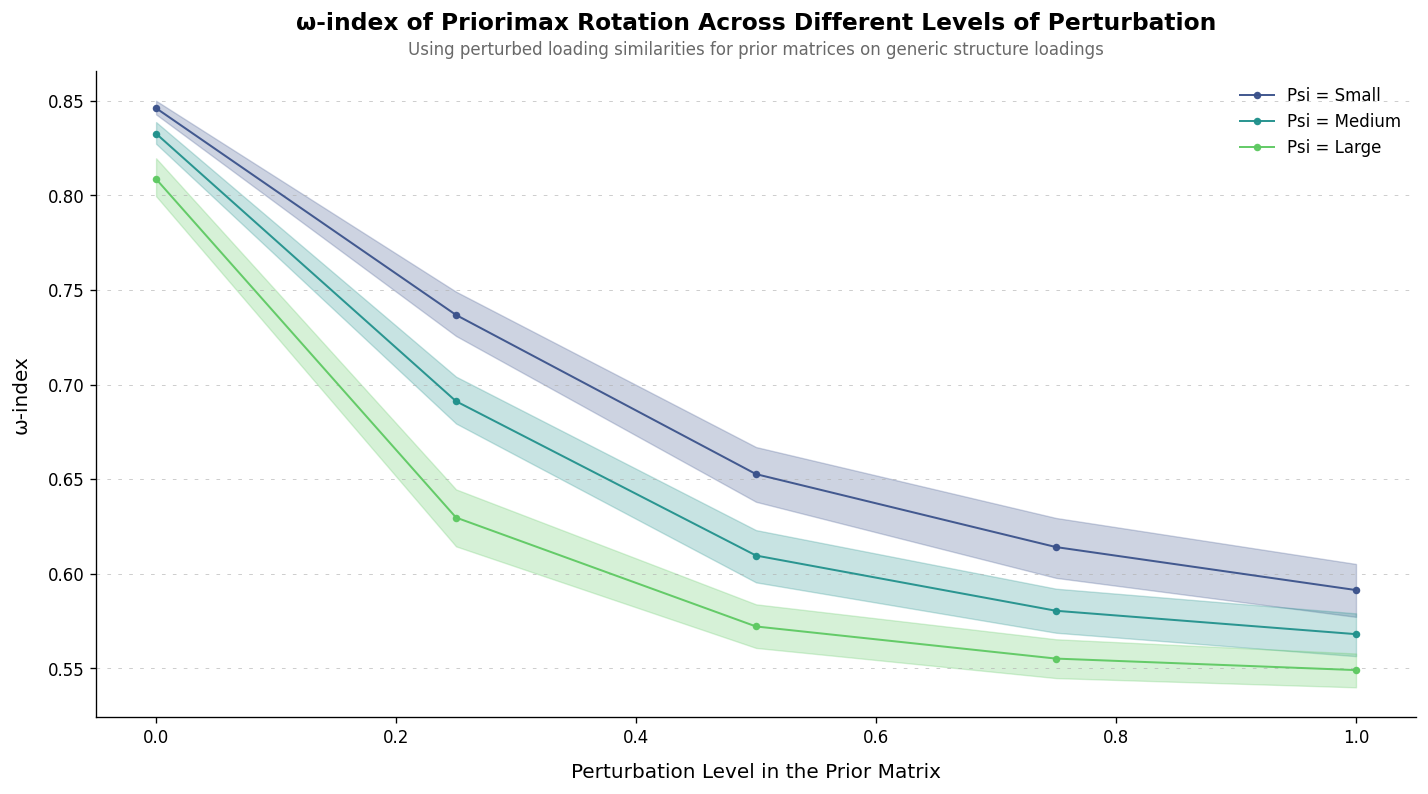}
      \caption{$\omega$ on Generic Loading Matrices\protect}
  \end{subfigure}
    \begin{subfigure}[c]{.5\textwidth}
      \centering
      \includegraphics[width=1\textwidth]{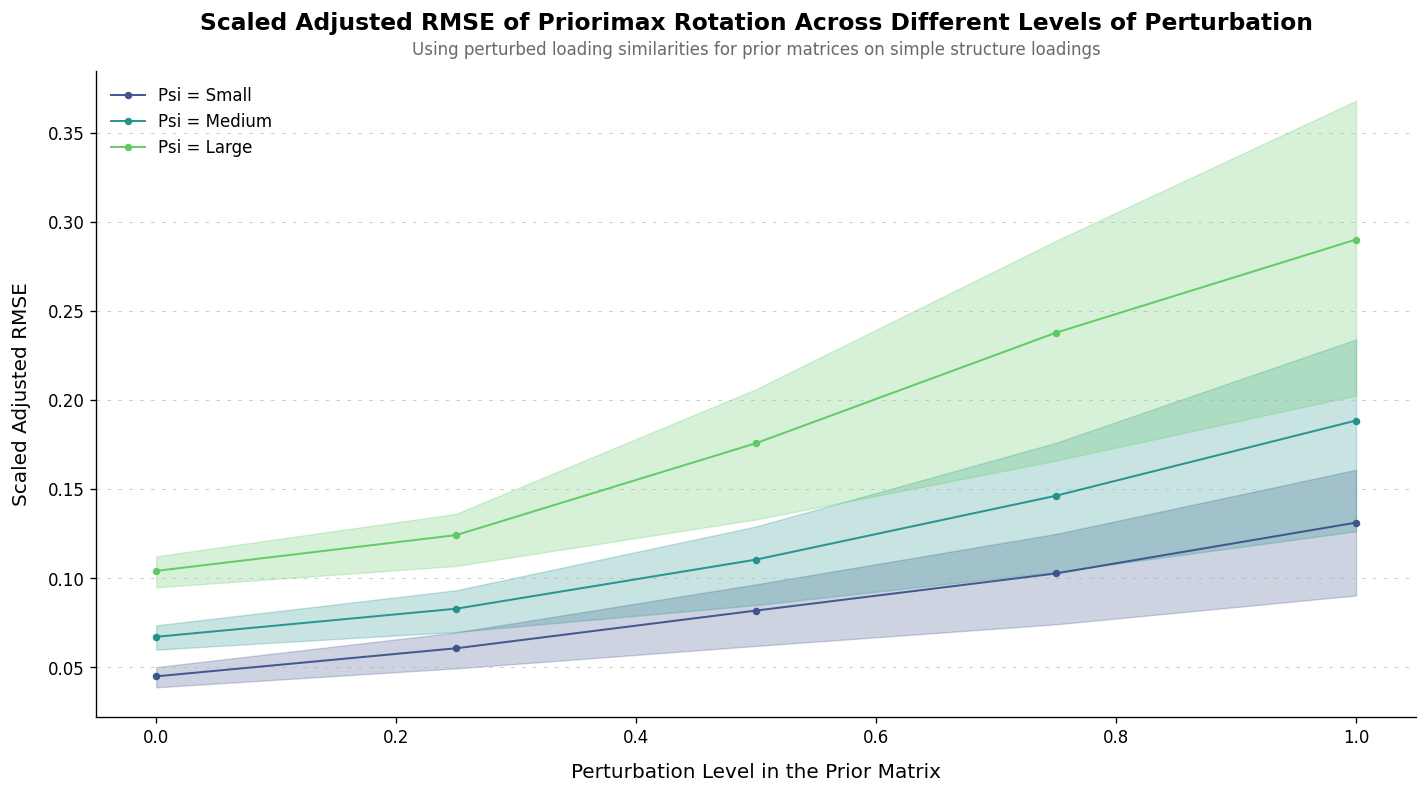}
      \caption{$\text{RMSE}_{\text{Scl.Adj.}}$ on Simple Loading Matrices}
  \end{subfigure}%
  \begin{subfigure}[c]{.5\textwidth}
      \centering
      \includegraphics[width=1\textwidth]{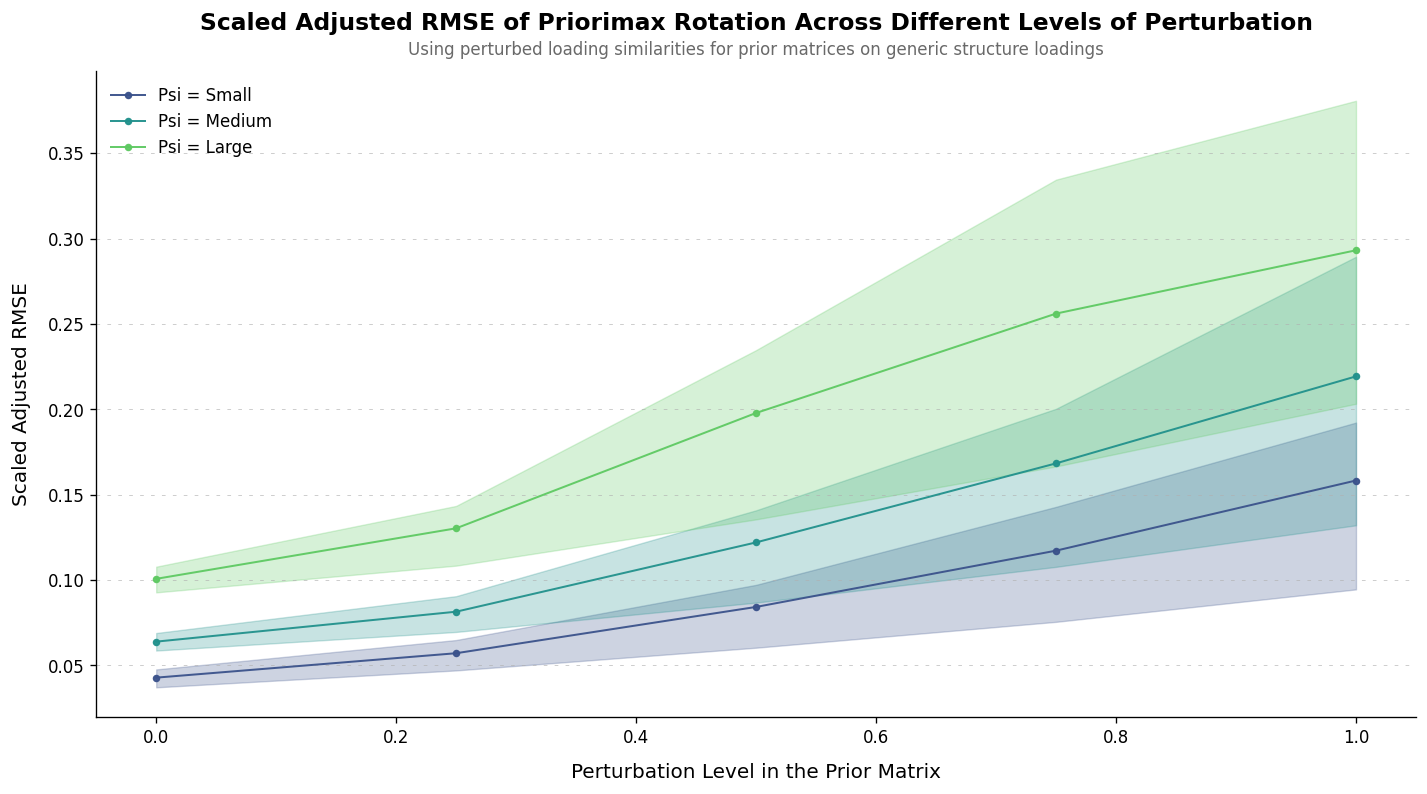}
      \caption{$\text{RMSE}_{\text{Scl.Adj.}}$ on Generic Loading Matrices\protect}
  \end{subfigure}
  \caption{Performance of Priorimax Rotation}
  \label{fig:priorimax-perf}
\end{figure}

First, let us look at the performance of the priorimax rotation on its own, in terms of both $\omega$ and $\text{RMSE}_{\text{Scl.Adj.}}$, as the communalities and level of perturbation in the prior matrix change. Refer to Figure \ref{fig:priorimax-perf}.
\vspace{-1.5em}
From Figure \ref{fig:priorimax-perf}, it can be seen that as the level of perturbation in the prior matrix increases, $\omega$ decreases. The same trend on $\omega$ applies as the specific variances increase (i.e., the communalities increase). These patterns appear to hold true for both simple and generic loading matrices. Furthermore, the performance of the priorimax rotation seems to be roughly the same for both types loading matrices in terms of $\omega$. On the other hand, the scaled adjusted RMSE increases as either the level of perturbation or the specific variances increase, for both types of loading matrices. Unlike in the cases for $\omega$ though, the scaled adjusted RMSE appears to become more variable as either specific variances or perturbation levels increase. Overall, the priorimax rotation performs better when perturbation levels and specific variances are small.

\begin{figure}[H]
  \centering
  \begin{subfigure}[c]{.5\textwidth}
      \centering
      \includegraphics[width=1\textwidth]{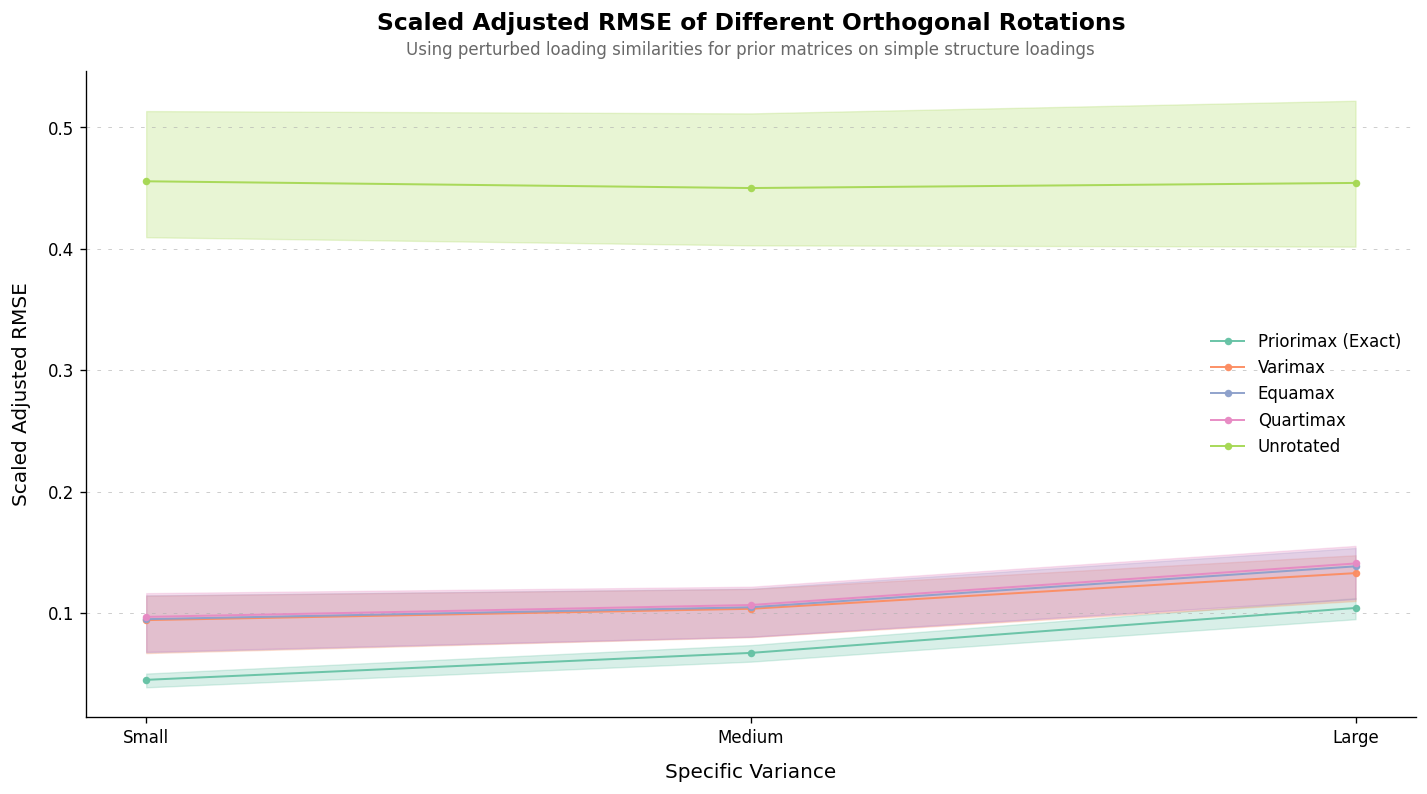}
      \caption{Simple Loading Matrices}
  \end{subfigure}%
  \begin{subfigure}[c]{.5\textwidth}
      \centering
      \includegraphics[width=1\textwidth]{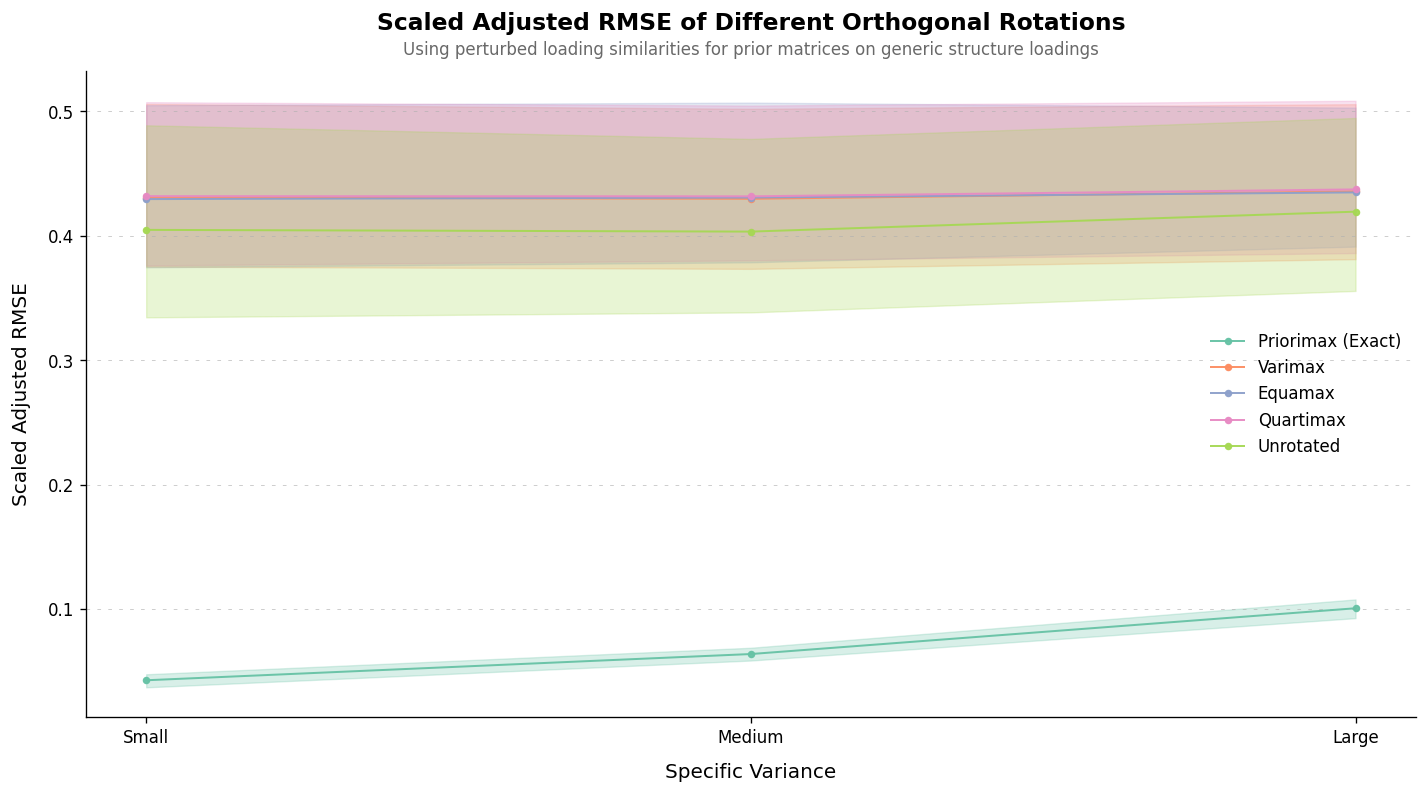}
      \caption{Generic Loading Matrices\protect}
  \end{subfigure}
  \caption{Comparison of Orthogonal Rotations}
  \caption*{{\footnotesize Note: For the priorimax rotation, it is assumed that there is no perturbation (i.e., exact prior matrix).}}
  \label{fig:priorimax-vs-others}
\end{figure}

Additionally, the results suggest that the interpretability of the factor model decreases as the communalities decrease. Ultimately, the findings from Figure \ref{fig:priorimax-perf} provide evidence that $\omega$ is indeed a suitable goodness-of-fit measure, as it indicates a better fit when noise levels are low, which is expected for any goodness-of-fit measure, similar to the case in \textcite{enhancedInterpretability2020}. This time, let us compare the priorimax rotation to other orthogonal rotations and refer to Figure \ref{fig:priorimax-vs-others}.
\vspace{-1.5em}
Looking at Figure \ref{fig:priorimax-vs-others}, the priorimax rotation outperforms all other orthogonal rotations (and the unrotated case) in terms of the scaled adjusted RMSE for both simple and generic loading matrices, but it shines even more when the true loading matrix does not follow a simple structure. Its performance is also much more stable compared to the other rotations. Although, note that the figure assumes that the prior matrix used is the true loading similarity matrix (i.e., no perturbation and inaccuracies). Combining these findings with Figure \ref{fig:priorimax-perf}, it appears that encoding information via the prior matrix can be useful not only for exploring interpretable configurations, but also for estimation itself, provided that the level of perturbation or ``noise" in the prior matrix is not too large. As a specific example, using the priorimax rotation with the semantic similarity matrix as the prior matrix can more closely and reliably recover the true loading matrix, especially when cross-loadings are present, if the semantics serve as good proxies for the loadings.

\begin{figure}[H]
  \centering
  \begin{subfigure}[c]{.5\textwidth}
      \centering
      \includegraphics[width=1\textwidth]{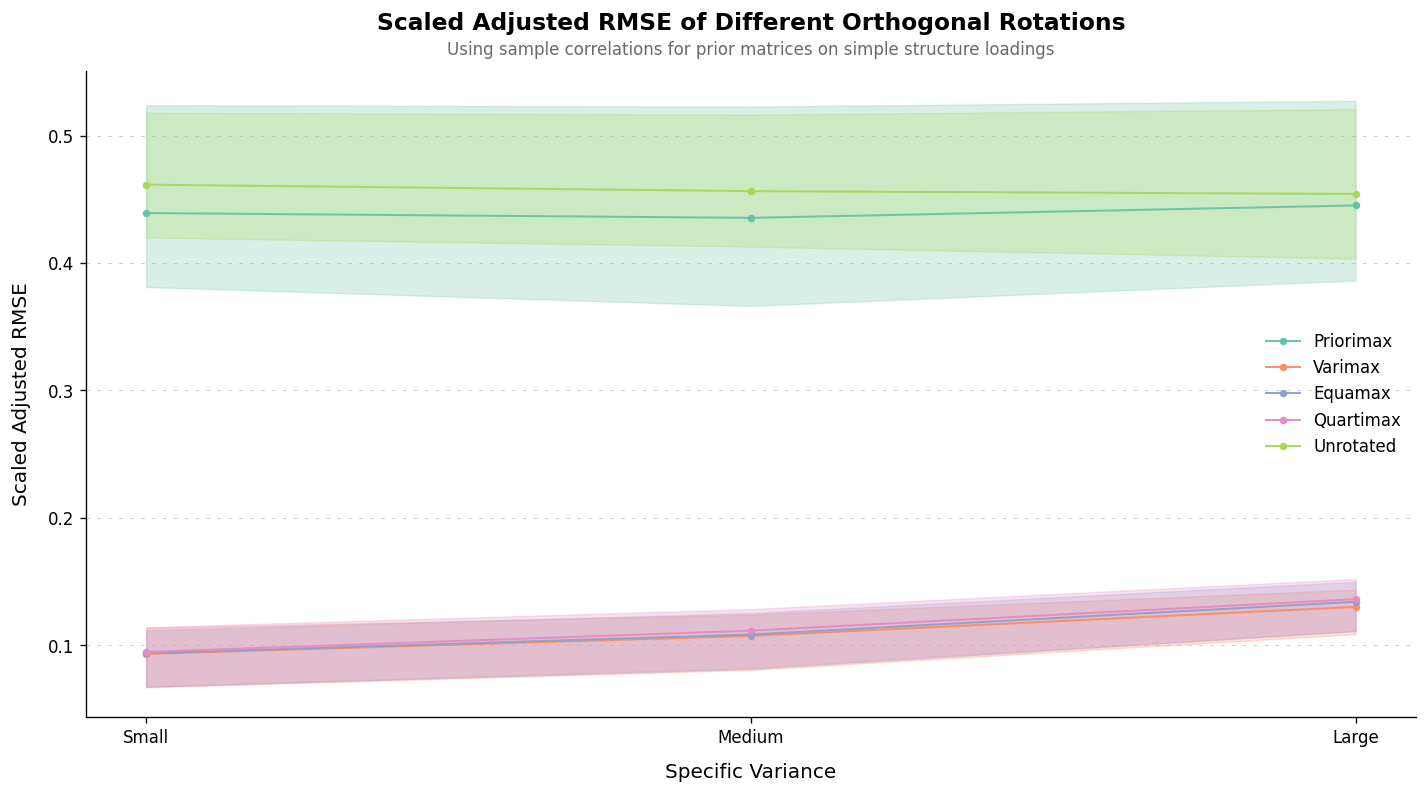}
      \caption{Simple Loading Matrices}
  \end{subfigure}%
  \begin{subfigure}[c]{.5\textwidth}
      \centering
      \includegraphics[width=1\textwidth]{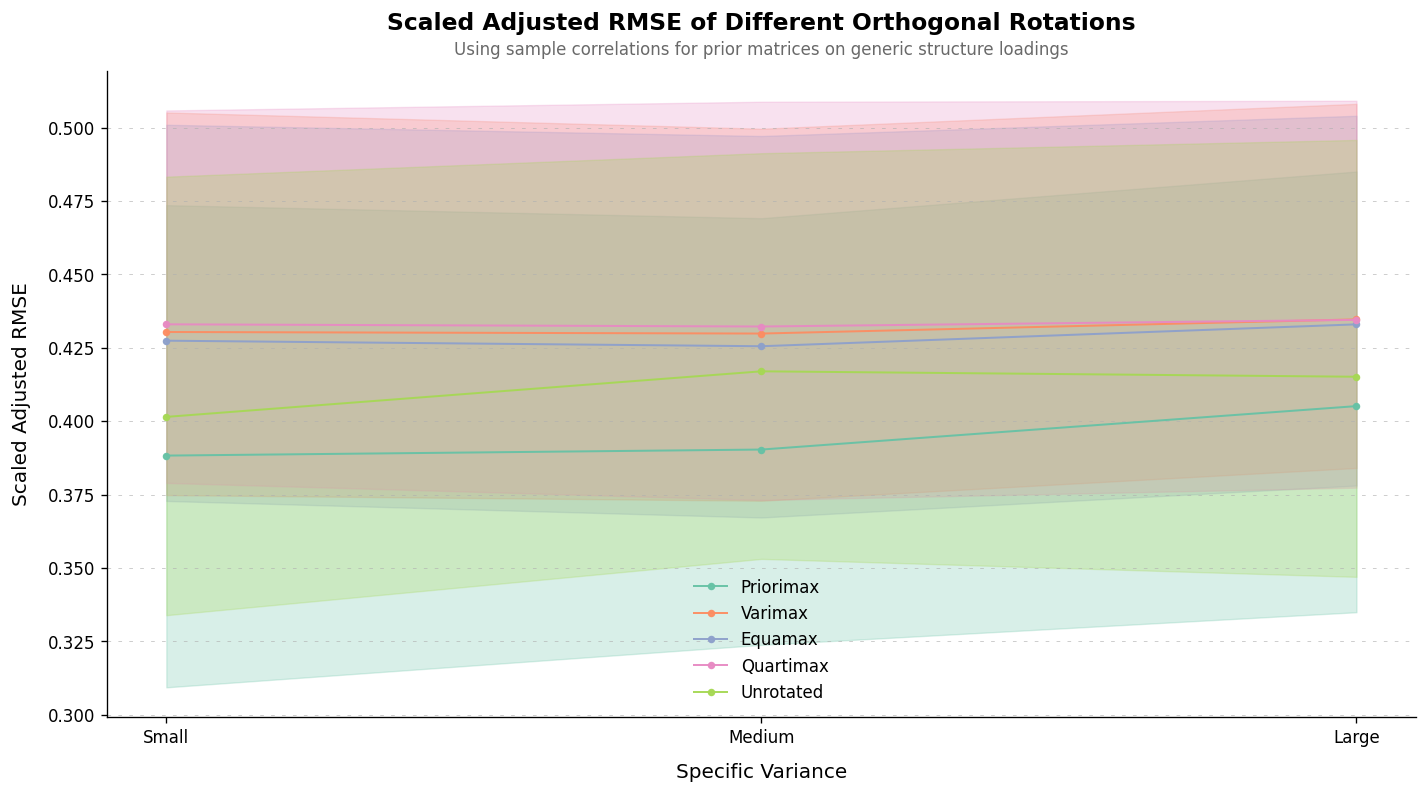}
      \caption{Generic Loading Matrices\protect}
  \end{subfigure}
  \caption{Using the Sample Correlation Matrix as the Prior Matrix}
  \label{fig:corr-as-prior}
\end{figure}

Finally, recall that the prior matrix is technically a similarity matrix and that the sample correlation matrix itself can be viewed as a similarity matrix. Thus, the researcher may consider using the sample correlation matrix as the prior matrix. In a sense, the sample correlation matrix is used both to estimate the unrotated loading matrix and to rotate the original loading matrix. To explore the viability of this approach, we extended the simulation study to consider cases where the prior matrix was the sample correlation matrix, as shown in Figure \ref{fig:corr-as-prior}. As seen in the figure, the priorimax rotation performs better compared to the unrotated case and to the other orthogonal rotations only if the true loading matrix does not follow a simple structure (e.g., when cross-loadings are present). The priorimax rotation actually performs worse for simple loading matrices, which is not surprising as the other orthogonal rotations generally seek a form of simple structure while the priorimax rotation does not. Hence, using the sample correlation matrix as the prior matrix may be useful if cross-loadings are expected.
\vspace{-1.5em}
\subsection{Empirical Demonstration}
To demonstrate the agreement between semantics and constructs, we applied the priorimax rotation to data obtained from established psychometric instruments with validated factor structures. Showing sufficient agreement between semantics and factors (particularly, semantic similarities and loading similarities) provides evidence linking language and latent constructs, supporting previous findings and the use of semantic similarities as ``proxies" for loading similarities in the exploratory context.

Now, the first instrument is the Depression Anxiety Stress Scales (DASS), developed by \textcite{lovibond}, a widely used $42$-item self-report questionnaire designed to measure three distinct yet closely related dimensions of negative emotional states: Depression, Anxiety, and Stress. In the clinical setting, DASS is primarily intended to identify the specific domain of emotional disturbance, facilitating a more precise clinical assessment. Respondents report their experiences during the previous week by rating each item on a four-point Likert scale ranging from $1$ (Did not apply to me at all) to $4$ (Applied to me very much, or most of the time).

Meanwhile, the second instrument is the Big Five Personality Test, which utilizes items from the Big-Five Factor Markers of the International Personality Item Pool (IPIP), developed by \textcite{bigFiveGoldberg}. The test comprises $50$ items that measure the five dimensions of personality, commonly referred to as the OCEAN model: Openness to experience, Conscientiousness, Extraversion, Agreeableness, and Neuroticism. Responses were recorded using a five-point Likert scale, ranging from $1$ (Disagree) to $5$ (Agree), with $3$ indicating a neutral response. Due to its robust psychometric properties, the test has been widely used in recruitment and organizational behavior research.

\begin{figure}[H]
  \centering
  \begin{subfigure}[c]{.5\textwidth}
      \centering
      \includegraphics[width=1\textwidth]{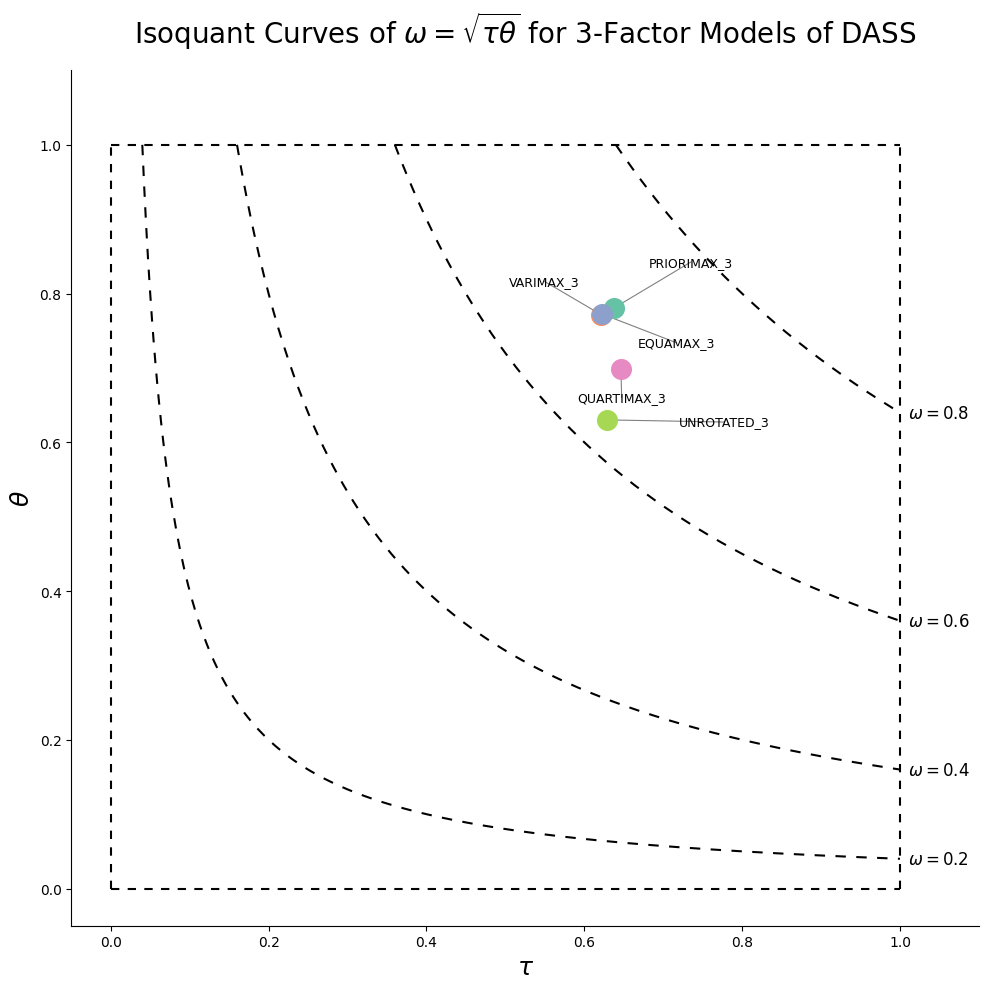}
      \caption{Using Depression Anxiety Stress Scales Dataset}
  \end{subfigure}%
  \begin{subfigure}[c]{.5\textwidth}
      \centering
      \includegraphics[width=1\textwidth]{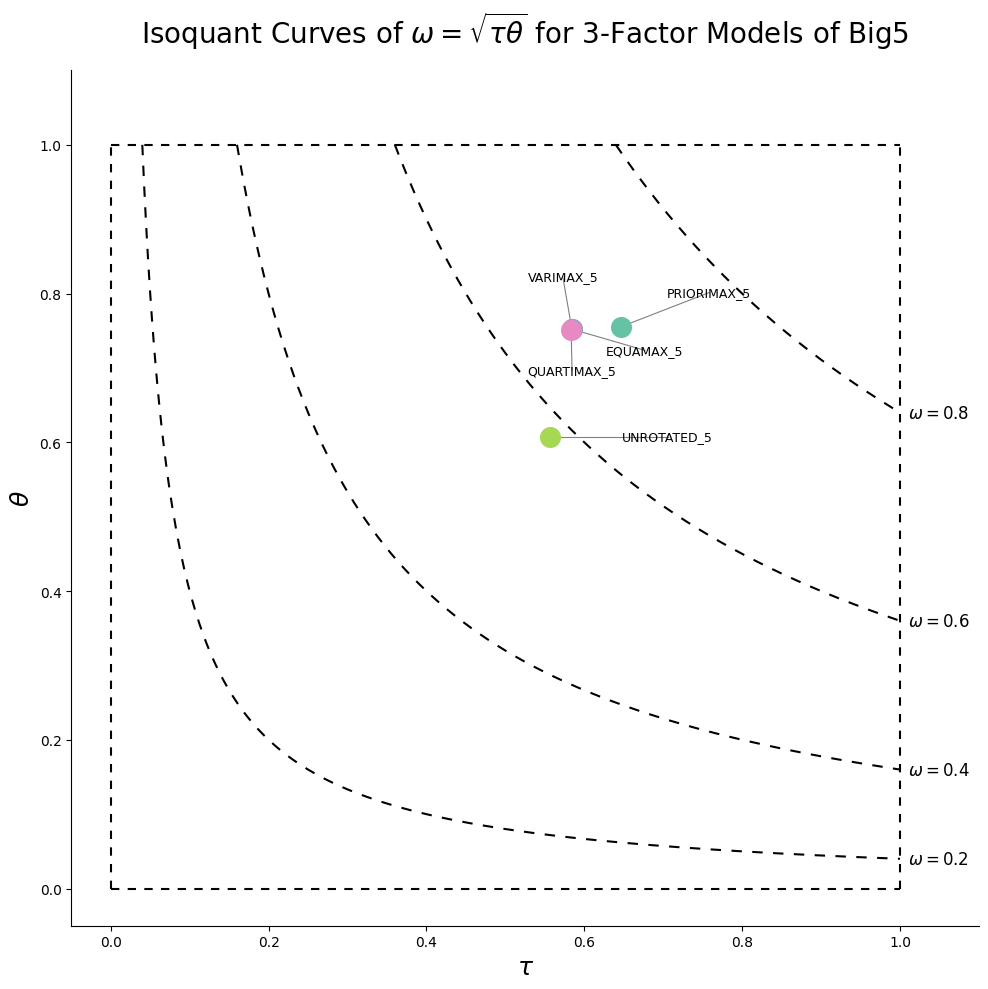}
      \caption{Using Big Five Personality Test Dataset\protect}
  \end{subfigure}
  \caption{Isoquant Plot for the Factor Models}
  \label{fig:isoquant-plot}
\end{figure}

Ultimately, the raw datasets for both instruments were obtained from \textcite{openpsych}, a publicly available online repository of psychological data. After data cleaning, the DASS dataset consisted of $39775$ sample points, whereas the Big Five dataset consisted of $19718$ sample points. Then, the priorimax rotation was done using semantic similarities for the prior matrix. In particular, the newly developed \codify{Python} package, \codify{interpretablefa}, was used to fit a factor model and apply the priorimax rotation. Shown in Figure \ref{fig:isoquant-plot} are the isoquant graphs generated for DASS and Big Five.

Applying the priorimax rotation to both instruments yields the highest $\omega$-index value of $0.7$. The corresponding $\tau$, $\theta$, and $\omega$ values are presented in Table \ref{table:summary_indices_DASS} and Table \ref{table:summary_indices_BIG}. Notably, in both cases, the unrotated model is substantially less meaningful compared to the rotated models, based on the $\omega$-index and its individual components. Moreover, looking at the isoquant plots, we see that all models exceed or are close to the $\omega=0.6$ isoquant curve, suggesting that there is already some agreement between semantics and loadings even before rotation. However, the choice of the rotation matrix still matters, as evidenced by the considerably spread out points in the graph.

\begin{table}[H]
  \centering
  \caption{Indices of 3-Factor Models Using DASS Dataset}
  {
  \begin{tabular}{cccc} \toprule
  Model & $\tau$ & $\theta$ & $\omega$ \\ \midrule
  $3$-Factor Model Priorimax & 0.637868 & 0.780783 & 0.705717 \\
  $3$-Factor Model Varimax & 0.621800 & 0.771362 & 0.692555 \\
  $3$-Factor Model Equamax & 0.622473 & 0.772532 & 0.693455 \\
  $3$-Factor Model Quartimax & 0.646314 & 0.698708 & 0.672001 \\
  $3$-Factor Model Unrotated & 0.629341 & 0.630117 & 0.629729 \\ \bottomrule
  \end{tabular}
  }
  \label{table:summary_indices_DASS}
\end{table}

\begin{table}[H]
  \centering
  \caption{Indices of 5-Factor Models Using Big Five Dataset}
  {
  \begin{tabular}{cccc} \toprule
  Model & $\tau$ & $\theta$ & $\omega$ \\ \midrule
  $5$-Factor Model Priorimax & 0.646897 & 0.755523 & 0.699103 \\
  $5$-Factor Model Varimax & 0.584269 & 0.752497 & 0.663069 \\
  $5$-Factor Model Equamax & 0.584177 & 0.752218 & 0.662894 \\
  $5$-Factor Model Quartimax & 0.583370 & 0.751092 & 0.661940 \\
  $5$-Factor Model Unrotated & 0.557149 & 0.607488 & 0.581774 \\ \bottomrule
  \end{tabular}
  }
  \label{table:summary_indices_BIG}
\end{table}

\begin{figure}[H]
  \centering
  \begin{subfigure}[c]{.5\textwidth}
      \centering
      \includegraphics[width=1\textwidth]{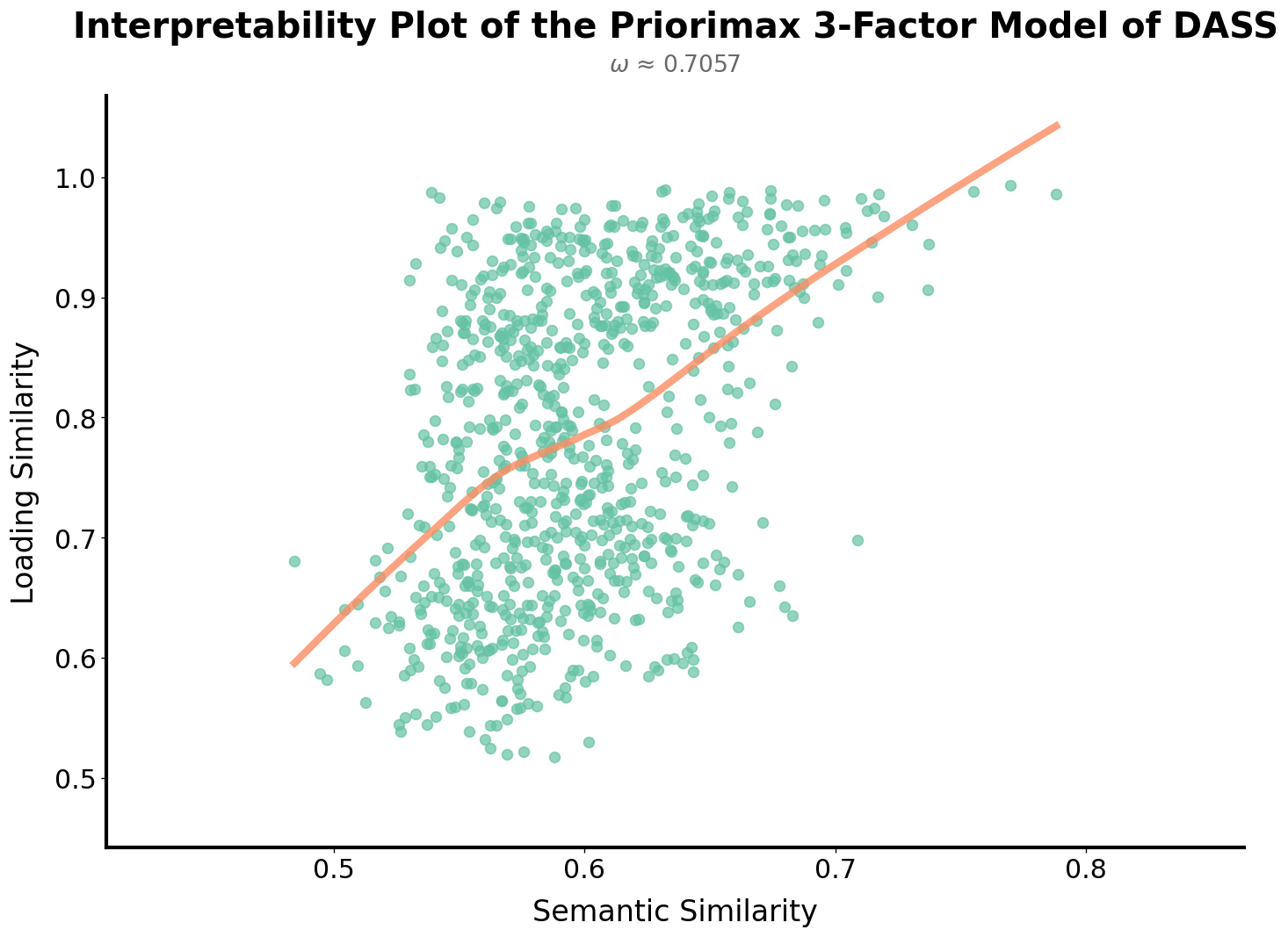}
      \caption{Using Depression Anxiety Stress Scales Dataset}
  \end{subfigure}%
  \begin{subfigure}[c]{.5\textwidth}
      \centering
      \includegraphics[width=1\textwidth]{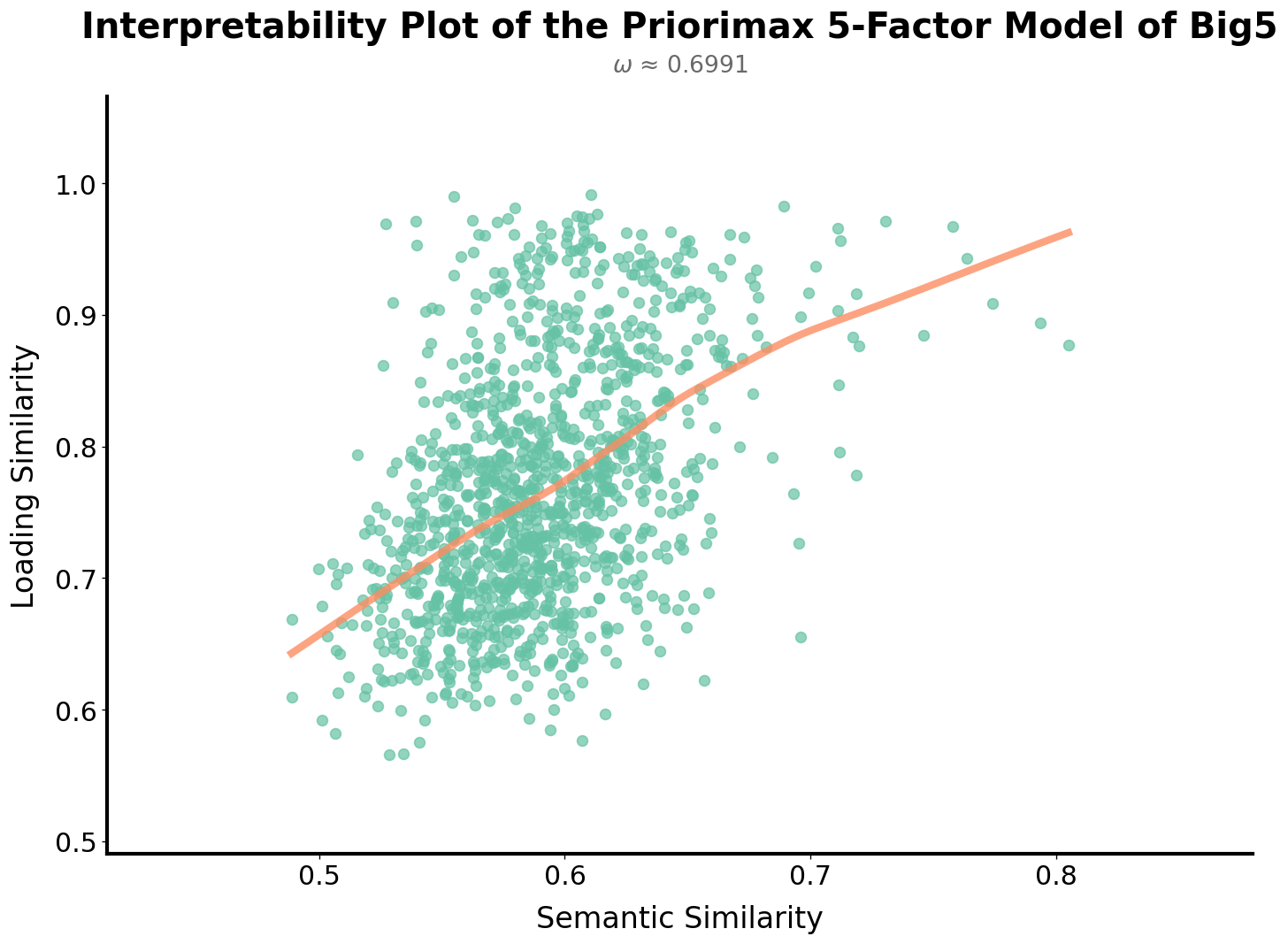}
      \caption{Using Big Five Personality Test Dataset\protect}
  \end{subfigure}
  \caption{Interpretability Plots for the Validated Number of Factors Using the Priorimax Rotation}
  \label{fig:interpretability_plot}
\end{figure}

Meanwhile, shown in Figure \ref{fig:interpretability_plot} are the interpretability plots for the priorimax rotations. It can be observed that for both datasets, there indeed appears to be a positive association between semantic similarity and loading similarity. For reference, the Pearson correlation coefficients for DASS and Big Five are about $0.42$ and $0.44$, respectively. When linearly mapped to $\left[0,1\right]$, the scale of $\omega$ and its components, these correspond to $0.71$ and $0.72$, respectively.

Ultimately, the $\omega$-index of $\approx 0.7$ obtained for both datasets provides empirical support for the relationship between semantics and psychological constructs, consistent with the findings of previous work. These findings reveal that a high $\omega$-index is achievable in practice.

\section{Conclusion}\label{sec:5}
In this paper, we propose a new interpretability index for factor models, $\omega$. The $\omega$-index serves as a goodness-of-fit measure that quantifies how ``interpretable" a factor model is by looking at how much the loading matrix aligns with the \textit{a priori} information. Based on $\omega$, we also created a factor rotation method, which we call pairwise target rotation or \textit{priorimax rotation}. Through the index and the rotation method, an intuitive yet flexible way of encoding \textit{a priori} information (including, but not limited to, semantic information) to influence the loading matrix towards a meaningful configuration is provided --- particularly useful for exploratory factor analysis.

Although, as the theoretical characterization suggested and as the simulation study showed, the priorimax rotation can also be useful for recovering the true loading matrix, especially when cross-loadings are present, provided that the prior matrix (e.g., the semantic similarity matrix) is a good proxy for the loadings. For instance, simply (re-)using the sample correlation matrix as the prior matrix may improve recoverability without the need to add more data or information. In the case of using semantic \textit{a priori} information, several studies \parencite{deepLexical2022,semanticLoadings,pfa2025,milano2025}, including the empirical case study in this paper, have demonstrated that the semantic similarity matrix can indeed possibly be a good proxy. Moreover, as argued in \textcite{pfa2025} and \textcite{milano2025}, examining the latent structure based on item semantics can possibly support questionnaire validation \textit{ex ante}. Thus, the interpretability index $\omega$ may also be used to aid validation \textit{ex post} (e.g., sufficiently large $\omega$ may indicate good validity).

In our approach, we introduce the concept of prior similarities and loading similarities, and work on the premise that the agreement between the two similarity matrices can aid in exploring interpretable configurations of factor models. One can view the sample correlation matrix, which is used to estimate the loading matrix up to a rotation, as a similarity matrix --- our approach then leverages another similarity matrix, the prior matrix, to resolve rotational indeterminacy. As opposed to other methods that aim for sparsity or utilize additional explanatory variables, our method is a modified version of target rotation, where the ``target" is specified as pairwise similarities as opposed to specific factor loadings and factor correlations (more intuitive in the case of semantic similarity, for instance).

With that being said, there are several extensions to our approach that can be explored further. One is to develop a \textit{weighted} version of $\omega$. The $\omega$-index essentially takes the geometric mean of two components, $\tau$ and $\theta$. However, there may be situations in which maximizing $\theta$ is considered more important than maximizing $\tau$. Thus, perhaps $\omega$ can be extended to accept a hyper-parameter $h\in\left[0,1\right]$ that controls the weights of the components, setting $\omega:=\tau^{h}\theta^{1-h}$ for example. Second, we considered only orthogonal rotations for the priorimax rotation in this paper --- extending the method to oblique rotations can be useful. Third, expanding the Monte Carlo experiment to include more experimental factors or conditions can also be helpful. Lastly, using other functions with the similarity-based and pairwise nature of the method may be interesting, as this study aims to encourage future research into the incorporation of semantics (or other forms of \textit{a priori} information) in factor analysis.

\section*{Acknowledgements}
The authors would like to thank Noeliano Ocorro and Rienyl Aiken Bacongallo for participating in brainstorming sessions during the early stages of the research study.

\printbibliography[heading=bibintoc]

\begin{appendices}

  \renewcommand{\addcontentsline}[3]{}

  \section{Loading Similarity}\label{appendix:A}
  First, observe that 
  $$
  0 \leq \tilde{l}_{i,j}^{2} \leq 1 \quad \text{and} \quad 0\leq\tilde{l}_{i,1}^{2}+\tilde{l}_{i,2}^{2}+\dots+\tilde{l}_{i,T}^{2}\leq 1
  $$
  Then, define $D_{f}:\left\{X_{1},X_{2},\dots,X_{M}\right\}\times\left\{X_{1},X_{2},\dots,X_{M}\right\}\to\mathbb{R}$ as
  $$
  D_{f}\left(X_{i},X_{j}\right)=\sqrt{\sum_{k=1}^{T}{\left(\tilde{l}_{i,k}^{2}-\tilde{l}_{j,k}^{2}\right)^{2}}}
  $$
  Note that $D_{f}\left(X_{i},X_{j}\right)$ is the Euclidean distance between the points $\left(\tilde{l}_{i,1}^{2},\tilde{l}_{i,2}^{2},\dots,\tilde{l}_{i,T}^{2}\right)$ and $\left(\tilde{l}_{j,1}^{2},\tilde{l}_{j,2}^{2},\dots,\tilde{l}_{j,T}^{2}\right)$ in the real $T$-space. Intuitively, $D_{f}$ measures how far $X_{i}$ and $X_{j}$ are in terms of their loadings. In order to make $D_{f}$ a similarity function and restrict its range to $\left[0,1\right]$ for ease of interpretability, observe that 
  $$
  \sum_{k=1}^{T}{\tilde{l}_{i,k}^{4}}\leq1-2\sum_{p=1}^{T-1}{\sum_{q=p}^{T}{\tilde{l}_{i,p}^{2}\tilde{l}_{i,q}^{2}}} \quad \text{and} \quad \sum_{k=1}^{T}{\tilde{l}_{j,k}^{4}}\leq1-2\sum_{p=1}^{T-1}{\sum_{q=p}^{T}{\tilde{l}_{j,p}^{2}\tilde{l}_{j,q}^{2}}}
  $$
  However, since $0 \leq \tilde{l}_{a,b}^{2} \leq 1$ for all $a$ and $b$, it follows that $\sum_{p=1}^{T-1}{\sum_{q=p}^{T}{\tilde{l}_{i,p}^{2}\tilde{l}_{i,q}^{2}}} \geq 0 $ and $\sum_{p=1}^{T-1}{\sum_{q=p}^{T}{\tilde{l}_{j,p}^{2}\tilde{l}_{j,q}^{2}}} \geq 0$. Consequently, $\sum_{k=1}^{T}{\tilde{l}_{i,k}^{4}}\leq 1$ and $\sum_{k=1}^{T}{\tilde{l}_{j,k}^{4}} \leq 1$. It also follows that $\sum_{k=1}^{T}{\tilde{l}_{i,k}^{2}\tilde{l}_{j,k}^{2}} \geq 0$.  Thus, 
  $$
  \sum_{k=1}^{T}{\left(\tilde{l}_{i,k}^{2}-\tilde{l}_{j,k}^{2}\right)^{2}}=\sum_{k=1}^{T}{\tilde{l}_{i,k}^{4}}+\sum_{k=1}^{T}{\tilde{l}_{j,k}^{4}}-2\sum_{k=1}^{T}{\tilde{l}_{i,k}^{2}\tilde{l}_{j,k}^{2}}\leq 1+1-2\cdot0=2
  $$
  The upper bound can be attained. For example, the quantity is exactly $2$ when $\left(\tilde{l}_{i,k_{1}}^{2},\tilde{l}_{j,k_{2}}^{2}\right)=\left(1,0\right)$ and $\left(\tilde{l}_{i,k_{2}}^{2},\tilde{l}_{j,k_{2}}^{2}\right)=\left(0,1\right)$ for some $k_{1}\neq k_{2}$. Thus,
  $$
D_{f}\left(X_{i},X_{j}\right)=1-\sqrt{\frac{1}{2}\sum_{k=1}^{T}{\left(\tilde{l}_{i,k}^{2}-\tilde{l}_{j,k}^{2}\right)^{2}}}\in\left[0,1\right]
$$

    \section{Simulating Loading Matrices}\label{appendix:B}
    \begin{algorithm}[H]
    \caption{Setting specific variances}
    \label{alg:setting-psis}
    \KwIn{Specific variance size $s \in \{\texttt{small}, \texttt{medium}, \texttt{large}\}$}
    \KwOut{Specific variances $\psi_j$ for manifest variables $j=1,\ldots,M$}
    
    \eIf{$s = \texttt{small}$}{
        \For{$j \leftarrow 1$ \KwTo $M$}{
            $\psi_j \sim U(0.1, 0.3)$\;
        }
    }{
        \eIf{$s = \texttt{medium}$}{
            \For{$j \leftarrow 1$ \KwTo $M$}{
                $\psi_j \sim U(0.3, 0.5)$\;
            }
        }{
            \For{$j \leftarrow 1$ \KwTo $M$}{
                $\psi_j \sim U(0.5, 0.7)$\;
            }
        }
    }
    \end{algorithm}
    
    \begin{algorithm}[H]
    \caption{Generating a simple loading matrix}
    \label{alg:generate-simple-loadings}
    
    \KwIn{Number of manifest variables $M$, number of factors $T$, specific variance size $s_{\psi}$}
    \KwOut{Loading matrix $\bm{\underline{L}}$}
    
    Generate specific variances $\psi_1,\ldots,\psi_M$ according to Algorithm~\ref{alg:setting-psis} with size $s_{\psi}$ as input\;
    
    Construct a simple structure indicator matrix $\bm{\underline{S}} \in \{0,1\}^{M \times T}$ by assigning each manifest variable to exactly one factor, such that each factor has the same number of manifest variables assigned to it\;
    
    Initialize $\bm{\underline{L}} \leftarrow \bm{\underline{0}}_{M \times T}$\;
    
    \For{$i \leftarrow 1$ \KwTo $M$}{
        Let $\psi_i$ be the specific variance for manifest variable $i$\;
        \For{$j \leftarrow 1$ \KwTo $T$}{
    
            Determine loading size $c_{i,j}$:
            \begin{itemize}
                \item \texttt{large}, if $S_{i,j} = 1$;
                \item \texttt{small}, if $S_{i,j} = 0$ and $U(0,1) > 0.25$;
                \item \texttt{medium}, otherwise.
            \end{itemize}
    
            Draw sign $s \in \{-1,+1\}$ with equal probability\;
    
            \uIf{$c_{i,j} = \texttt{large}$}{
                Draw $u \sim U(0.8, 1.0)$\;
            }
            \uElseIf{$c_{i,j} = \texttt{medium}$}{
                Draw $u \sim U(0.4, 0.6)$\;
            }
            \Else{
                Draw $u \sim U(0, 0.2)$\;
            }
    
            Set $l_{i,j} \leftarrow s \cdot u$\;
        }
    
        Normalize row $i$:
        \[
        \bm{\underline{L}}_{i, \cdot}
        \leftarrow
        \frac{\bm{\underline{L}}_{i, \cdot}}
             {\sqrt{\sum_{j=1}^{T} l_{i,j}^{2}}}
        \sqrt{1 - \psi_i},
        \]
    }
    
    \Return{$\bm{\underline{L}}$\;}
    
    \end{algorithm}

    \begin{algorithm}[H]
    \caption{Generating a generic loading matrix}
    \label{alg:generate-generic-loadings}
    
    \KwIn{Number of manifest variables $M$, number of factors $T$, specific variance size $s_{\psi}$}
    \KwOut{Loading matrix $\bm{\underline{L}}$}
    
    Generate specific variances $\psi_1,\ldots,\psi_M$ according to Algorithm~\ref{alg:setting-psis} with size $s_{\psi}$ as input\;
    
    Initialize $\bm{\underline{L}} \leftarrow \bm{\underline{0}}_{M \times T}$\;
    
    \For{$i \leftarrow 1$ \KwTo $M$}{
        
        Let $\psi_i$ be the specific variance for manifest variable $i$\;
    
        \For{$j \leftarrow 1$ \KwTo $T$}{
    
            Assign loading size:
            \[
            c_{i,j} \sim \text{Uniform}\{\texttt{large}, \texttt{medium}, \texttt{small}\}
            \]\;
    
            Draw sign $s \in \{-1, +1\}$ with equal probability\;
    
            \uIf{$c_{i,j} = \texttt{large}$}{
                Draw $u \sim U(0.8, 1.0)$\;
            }
            \uElseIf{$c_{i,j} = \texttt{medium}$}{
                Draw $u \sim U(0.4, 0.6)$\;
            }
            \Else{
                Draw $u \sim U(0.0, 0.2)$\;
            }
    
            Set $l_{i,j} \leftarrow s \cdot u$\;
        }
    
        Normalize row $i$:
        \[
        \bm{\underline{L}}_{i, \cdot}
        \leftarrow
        \frac{\bm{\underline{L}}_{i, \cdot}}
             {\sqrt{\sum_{j=1}^{T} l_{i,j}^{2}}}
        \sqrt{1 - \psi_i},
        \]
    }
    
    \Return{$\bm{\underline{L}}$\;}

    \end{algorithm}

    \section{Generating Synthetic Data}\label{appendix:C}
    \begin{algorithm}[H]
    \caption{Generating synthetic data from a factor model}
    \label{alg:generate-data}
    
    \KwIn{Loading matrix $\bm{\underline{L}} \in \mathbb{R}^{M \times T}$, specific variances $\psi_1,\ldots,\psi_M$, sample size $N$}
    \KwOut{Data matrix $\bm{\underline{D}} \in \mathbb{R}^{N \times M}$}
    
    Initialize $\bm{\underline{D}} \leftarrow \bm{\underline{0}}_{N \times M}$\;
    
    \For{$n \leftarrow 1$ \KwTo $N$}{

        Draw factor scores by taking a random sample $f_{1},f_{2},\dots,f_{T}$ from $\mathcal{N}\left(0,1\right)$;
    
        \For{$i \leftarrow 1$ \KwTo $M$}{
    
            Compute for the contribution of common factors:
            \[
            s_{n,i} \leftarrow \sum_{j=1}^{T}{l_{i,j}f_{j}}
            \]\;
    
            Draw for the contribution of the specific factor:
            \[
            \varepsilon_{n,i} \sim \mathcal{N}(0, \psi_i)
            \]\;
    
            Set:
            \[
            d_{n,i} \leftarrow s_{n,i} + \varepsilon_{n,i}
            \]\;
        }
    }
    
    \Return{$\bm{\underline{D}}$}\;

    \end{algorithm}

    \section{Simulating Prior Matrices}\label{appendix:D}
    \begin{algorithm}[H]
    \caption{Generating a prior matrix}
    \label{alg:generating-prior}
    
    \KwIn{Loading similarity matrix $\bm{\underline{S}} \in \mathbb{R}^{M \times M}$, perturbation level $\delta$}
    \KwOut{Perturbed similarity matrix $\bm{\underline{S}}^{*}$}
    
    Initialize perturbation matrix $\bm{\underline{P}} \leftarrow \bm{\underline{0}}_{M \times M}$\;
    
    \For{$i \leftarrow 1$ \KwTo $M$}{
        \For{$j \leftarrow i+1$ \KwTo $M$}{
            
            Draw $\eta_{i,j} \sim U(-\delta, \delta)$\;
    
            Set $p_{i,j} \leftarrow \eta_{i,j}$\;
            Set $p_{j,i} \leftarrow p_{i,j}$\;
        }
    }
    
    Set $\bm{\underline{S}}^{*} \leftarrow \bm{\underline{S}} + \bm{\underline{P}}$\;
    
    \If{$\min(\bm{\underline{S}}^{*}) < 0$}{
        Shift matrix:
        \[
        \bm{\underline{S}}^{*} \leftarrow \bm{\underline{S}}^{*} + \left|\min(\bm{\underline{S}}^{*})\right|
        \]
    }
    
    \If{$\max(\bm{\underline{S}}^{*}) > 1$}{
        Rescale matrix:
        \[
        \bm{\underline{S}}^{*} \leftarrow \frac{\bm{\underline{S}}^{*}}{\max(\bm{\underline{S}}^{*})}
        \]
    }
    
    Set diagonal:
    \[
    s^{*}_{i,i} \leftarrow 1 \quad \forall i = 1,\ldots,M
    \]
    
    \Return{$\bm{\underline{S}}^{*}$}\;
    \end{algorithm}
\end{appendices}

\end{document}